\documentclass{aa}

\usepackage[varg]{txfonts}
\usepackage{longtable}
\usepackage{bm}

\newcommand\kms{km~s$^{-1}$}
\newcommand\msun{$M_\odot$}
\newcommand\mhi{$M_{\mathrm{HI}}$}
\newcommand\mstar{$M_{*}$}

\newcommand\vlsr{$v_{LSR}$}

\def\be{\begin{equation}}
\def\ee{\end{equation}}
\def\a40{$\alpha$.40}
\def\arcmin{$^{\prime}$}
\def\arcsec{$^{\prime\prime}$}
\def\dg{$^{\circ}$}

\newcommand{\hi}{H\,{\sc i}}
\def\mdyn{$M_{dyn}$}
\def\w50{$W_{50}$}

\defcitealias{2013ApJ...768...77A}{A13}
\defcitealias{2015ApJ...800L..15B}{B15}
\defcitealias{2015arXiv150300720S}{S15}

\usepackage{natbib,twoopt}
\usepackage[breaklinks=true]{hyperref} 
\bibpunct{(}{)}{;}{a}{}{,} 
\makeatletter
\newcommandtwoopt{\citeads}[3][][]{\href{http://adsabs.harvard.edu/abs/#3}%
{\def\hyper@linkstart##1##2{}%
\let\hyper@linkend\@empty\citealp[#1][#2]{#3}}}
\newcommandtwoopt{\citepads}[3][][]{\href{http://adsabs.harvard.edu/abs/#3}%
{\def\hyper@linkstart##1##2{}%
\let\hyper@linkend\@empty\citep[#1][#2]{#3}}}
\newcommandtwoopt{\citetads}[3][][]{\href{http://adsabs.harvard.edu/abs/#3}%
{\def\hyper@linkstart##1##2{}%
\let\hyper@linkend\@empty\citet[#1][#2]{#3}}}
\newcommandtwoopt{\citeyearads}[3][][]%
{\href{http://adsabs.harvard.edu/abs/#3}
{\def\hyper@linkstart##1##2{}%
\let\hyper@linkend\@empty\citeyear[#1][#2]{#3}}}
\makeatother

\begin{document}

\title{Identifying galaxy candidates in WSRT \hi\ imaging of
ultra-compact high velocity clouds}

\author{
Elizabeth A. K. Adams
\inst{1}
\and
Tom A. Oosterloo
\inst{1,2}
\and
John M. Cannon
\inst{3}
\and
Riccardo Giovanelli
\inst{4}
\and
Martha P. Haynes
\inst{4}
}
\institute{
ASTRON, Netherlands Institute for Radio Astronomy, Postbus 2, 7900 AA Dwingeloo, The Netherlands
 		\email{adams@astron.nl}
		\and
		Kapteyn Astronomical Institute, University of Groningen Postbus 800, 9700 AV Groningen, the Netherlands
		\and
		Department of Physics and Astronomy, Macalaster College, 1600 Grand Avenue, Saint Paul, MN 55105, USA
		\and
		Center for Astrophysics and Planetary Science, Space Sciences Building, Cornell University, Ithaca, NY 14853, USA
}

\abstract{
Ultra-compact high velocity clouds (UCHVCs) were identified in the 
Arecibo Legacy Fast ALFA (ALFALFA) \hi\ survey as 
  potential gas-bearing dark matter halos.
Here we present higher resolution neutral hydrogen (\hi) observations of twelve UCHVCS
 with the Westerbork Synthesis Radio Telescope (WSRT).
The UCHVCs were selected based on a combination of size, isolation, large recessional velocity and high column density 
as the best candidate dark matter halos.
The WSRT data were tapered to image the UCHVCs
 at 210\arcsec\ (comparable to the Arecibo resolution) and 105\arcsec\ angular resolution. 
 In a comparison of the single-dish to interferometer data, we find that the integrated line flux recovered in the WSRT
 observations is generally comparable to that from the single-dish ALFALFA data.
 In addition, any structure seen in the ALFALFA data is reproduced in the WSRT maps at the same angular resolution.
 At 210\arcsec\ resolution all the sources are generally compact with a smooth \hi\ morphology, as expected
 from their identification as UCHVCs. 
 At the higher angular resolution, 
 a majority of the sources break into small clumps contained in a diffuse envelope.
 These UCHVCs 
  also have no ordered velocity motion and
 are most likely Galactic halo clouds.
 We identify two UCHVCs, AGC\,198606 and AGC\,249525,
 as excellent galaxy candidates based on maintaining a smooth \hi\
 morphology at higher angular resolution and showing ordered velocity motion
 consistent with rotation.
 A third source, AGC\,249565, lies between these two populations in 
 properties and is a possible galaxy candidate.
If interpreted as gas-bearing dark matter halos,
the three candidate galaxies have rotation velocities of $8-15$ \kms,
\hi\ masses of $0.6-50 \times 10^{5}$ \msun, \hi\ radii of $0.3 - 2$ kpc,
and dynamical masses of $2-20 \times 10^7$ \msun\
for a range of plausible distances.
These are the UCHVCs with the highest column density values in the ALFALFA \hi\ data and we suggest this is the best way to
identify further candidates.
}


\titlerunning{Identifying Galaxy Candidates in WSRT \hi\ Imaging of UCHVCs}

\maketitle

\section{Introduction}

Studying the properties of the smallest galaxies
is important both for testing $\Lambda$CDM and 
 understanding galaxy formation.
$\Lambda$CDM does an excellent job of describing 
large scale structure and the distribution of massive galaxies \citepads[e.g.,][]{2014Natur.509..177V}.
However, on small scales, there are tensions between observations of dwarf galaxies
and predictions from simulations for both field dwarf galaxies and satellite dwarf galaxies
in the Milky Way (MW).
These tensions include the total number count of expected galaxies 
\citepads[e.g., the "missing satellites problem" in the Local Group;][] {1993MNRAS.264..201K,1999ApJ...522...82K,1999ApJ...524L..19M,2010ApJ...723.1359M,2011ApJ...739...38P};
which dark matter halos host galaxies 
 \citepads[e.g., the "too big too fail" problem;][]{2012MNRAS.422.1203B,2015A&A...574A.113P};
 and the structure of those dark matter halos \citepads[e.g., the "cusp-core" problem;][]{2008AJ....136.2648D,2011ApJ...742...20W}. 
Much work exists to suggest that these differences can be reconciled by the proper inclusion of
baryonic physics in simulations 
\citepads[e.g.,][]{2011AJ....142...24O,2015MNRAS.454.2092O,2016MNRAS.457.1931S,2016arXiv160205957W}.
However,
much of the relevant physics is implemented at the sub-grid level, and simulations
of dwarf galaxies are sensitive to (at least some of) these parameters 
\citepads[e.g.,][]{2010ApJ...710..408B,2013MNRAS.432.1989S,2016MNRAS.458..912V},
and
results between various simulations 
do not necessarily agree
\citepads[e.g., the existence of a \mstar-$M_{halo}$ relation at low masses,][]{2015MNRAS.448.2941S,2015MNRAS.454.2092O,2015MNRAS.453.1305W}.
Part of the issue may be resolution; as the resolution of simulations increases, typically the lowest
mass galaxy that can form in the simulation also decreases \citepads{2006MNRAS.371..401H,2015MNRAS.453.1305W}.

The existence of extremely low-mass galaxies that have a significant reservoir
of neutral hydrogen (\hi)
also
challenges our understanding of how the smallest galaxies form and evolve.
How have these galaxies maintained their gas reservoir given the multitude of 
baryonic processes that can disrupt it?
The most extreme example is Leo T;
this galaxy has an \hi\ mass of only $2.8 \times 10^5$ \msun\
and a stellar mass of $1.05 \times 10^5$ \msun\
\citepads{2008MNRAS.384..535R,2012ApJ...748...88W}.
The recent discovery of Leo P
demonstrates that there are more low-mass gas-dominated systems
to be discovered.
Leo P has an \hi\ mass of $8.1 \times 10^5$ and a stellar mass of
$5.6 \times 10^5$ \msun\ \citepads{2015ApJ...812..158M}.
Both systems are very faint optically;
Leo T is on the edge of detection for 
 the Sloan Digital Sky Survey
\citepads[SDSS,][]{2010AdAst2010E...8K},
and Leo P was originally identified as an \hi\ source
\citepads{2013AJ....146...15G}.
This highlights a parameter space for \hi\ surveys for understanding the smallest galaxies.
Galaxies similar to Leo T or Leo P but that lie at larger distances or have
had slightly different star formation histories (e.g., less recent star formation)
might be missed by optical surveys but detected in blind \hi\ surveys.

To this end, \citetads{2010ApJ...708L..22G} presented the idea that ultra-compact high
velocity clouds
(UCHVCs) in the Arecibo Legacy Fast ALFA (ALFALFA) \hi\ survey may be gas in dark matter halos
with a stellar counterpart not detectable in extant optical surveys.
\citetads[][hereafter A13]{2013ApJ...768...77A} built on this work, 
presenting a catalog of sources with specific selection criteria for the 40\% complete ALFALFA survey. 
Similarly, \citetads{2012ApJ...758...44S} presented a catalog of compact \hi\ clouds from GALFA \hi\ survey,
highlighting those that were potentially good galaxy candidates. 
These catalogs represent an excellent starting point for identifying potential gas-bearing dark matter halos in 
the local universe,
 and theoretical models suggest they contain good candidates \citepads[e.g.,][]{2013ApJ...777..119F}.
However, without stellar counterparts there are no direct distances to these clouds
and they may be local clouds of gas that arise from various Galactic processes.
Determining which single-dish \hi\ properties of these clouds are the best predictor 
that a system is a good candidate to be 
gas in a dark matter halo, rather than a local \hi\ cloud,  is critical for making the most use of expensive follow-up observations.

The most straightforward approach for understanding the nature of the UCHVCs would be to detect
an optical counterpart, constraining the distance to the system and identifying it as a bona-fide galaxy.
Previous work has shown that stellar counterparts for UCHVCs appear to be rare 
and correspond to more distant, massive systems with typical
\hi\ masses of $\sim 10^7$ \msun\
\citepads{2015A&A...575A.126B,2015ApJ...806...95S}.
However, \citetads{2015ApJ...811...35J} identified a tentative optical counterpart for one UCHVC,
AGC\,198606,
at a distance of 383 kpc. This distance is consistent with the hypothesis that this UCHVC is a companion to Leo T,
located only 1.2\dg\ and 17 \kms\ away.
At this distance, AGC\, 198606 has an \hi\ mass of $5\times10^5$ \msun,
comparable to that of Leo T, while its 
stellar mass is an order of magnitude lower. 

The lack of definitive stellar counterparts to date may indicate that UCHVCs
representing gas-bearing dark matter halos are rare. Alternatively, it could
also be a result of these systems having intrinsically faint stellar counterparts;
in the case above, AGC\,198606 has  \mhi/\mstar $>40$.
The ability to better distinguish the best candidates could help address
which of these scenarios is dominant.
Previous work with high velocity clouds (HVCs), and especially compact HVCs 
(previously proposed as "dark" galaxies),
has shown that higher resolution \hi\ imaging can be used to constrain the nature of these systems and
to address whether they are Galactic or extragalactic \citepads[e.g.,][]{2002A&A...391...67D,2004A&A...426L...9B,2005A&A...436..101W,2014A&A...563A..99F}.

Hence we present resolved \hi\ observations with the Westerbork Synthesis Radio Telescope (WSRT) data for
 twelve UCHVCs in order to help address their nature.
The sources are drawn from a catalog of UCHVCs
following the selection criteria of 
\citetalias{2013ApJ...768...77A} but including expanded sky coverage of
the ALFALFA \hi\ survey.
Importantly, the restriction that $| v_{LSR} | > 120$ \kms\ was relaxed so that 
clouds close to Galactic \hi\ velocities that were otherwise good candidates are included.
The sources were selected to represent the best potential galaxy candidates on the
basis of various properties: 
high average column density, as those systems with the highest density of gas
and potential for star formation; 
small angular size, for the systems most
consistent with being distant objects; 
isolation, as the objects least likely to be part of a larger Galactic HVC complex;
and large recessional velocity, as it is difficult to explain in a Galactic fountain model.
Every cloud was selected for at least one of these criterion; a few fulfilled multiple criteria.
The goal is to determine which, if any, of these criteria
 are most important
for identifying the best candidate galaxies.

In Section 2 we present the ALFALFA \hi\ properties and the WSRT \hi\ data for the twelve UCHVCs observed.
In Section 3, we compare the ALFALFA and WSRT \hi\ properties.
 In Section 4 we discuss the nature of the UCHVCs, how to identify the best galaxy candidates,
 and the properties of the galaxy candidates.
We summarize
our results in Section 5.
The Appendix contains a full presentation of the data products for all the UCHVCs.

\begin{table*}
\footnotesize
\caption{ALFALFA \hi\ properties}
\label{tab:hi_alfalfa}
\centering
\begin{tabular}{lllllllll}
\hline \hline
HVC name & AGC & R.A.+Dec. & cz & \w50 & $\bar a$ & $S_{HI}$ & $M_{HI}$ & $\log \bar N_{HI}$ \\
          &    &  J2000    & \kms & \kms & \arcmin & Jy \kms & \msun    & atoms cm$^{-2}$ \\
(1) & (2) & (3) & (4) & (5) & (6) & (7) & (8) &(9)\\
\hline
HVC214.76+42.45+44 & 198606\tablefootmark{1,2} & 093005.4+163956 & 51 & 26 (1) & 12.2 & 17.44 (0.05) & 6.61 & 19.71\\
HVC204.88+44.86+147 & 198511\tablefootmark{2,3} & 093013.2+241217 & 152 & 15 (1) & 7.0 & 0.73 (0.03) & 5.24 & 18.81\\
HVC205.83+45.14+173 & 198683 & 093208.0+233752 & 178 & 19 (1) & 10.4 & 0.88 (0.04) & 5.32 & 18.56\\
HVC217.77+58.67+96 & 208747\tablefootmark{2} & 103706.6+203058 & 98 & 23 (1) & 10.9 & 2.74 (0.05) & 5.81 & 19.0\\
HVC212.68+62.39+64 & 208753 & 104932.4+235638 & 65 & 23 (1) & 12.7 & 3.95 (0.07) & 5.97 & 19.03\\
HVC230.27+71.10+76 & 219663 & 113429.7+201249 & 74 & 17 (1) & 7.2 & 0.75 (0.03) & 5.25 & 18.8\\
HVC235.38+74.79+195 & 219656\tablefootmark{2} & 115124.3+203220 & 192 & 21 (1) & 7.5 & 0.88 (0.04) & 5.32 & 18.84\\
HVC271.57+79.03+248 & 229326\tablefootmark{2} & 122734.7+173823 & 242 & 23 (8) & 7.8 & 0.77 (0.04) & 5.26 & 18.74\\
HVC276.53+79.84+255 & 229327\tablefootmark{2} & 123231.6+175721 & 249 & 19 (1) & 11.1 & 0.9 (0.05) & 5.33 & 18.51\\
HVC028.09+71.87-142 & 249393\tablefootmark{2,3} & 141054.9+241210 & -155 & 38 (2) & 12.7 & 1.02 (0.07) & 5.38 & 18.44\\
HVC011.76+67.89+60 & 249525\tablefootmark{2} & 141750.1+173252 & 48 & 24 (7) & 8.5 & 6.36 (0.04) & 6.18 & 19.59\\
HVC015.96+63.90+44 & 249565 & 143557.6+171004 & 30 & 18 (1) & 7.5 & 1.76 (0.04) & 5.62 & 19.14\\
\hline
\end{tabular}
\tablefoot{
\tablefootmark{1}{Previously published in  \citetads{2015A&A...573L...3A}.}
\tablefootmark{2}{Previously published in \citetads{2016MNRAS.457.4393J}.}
\tablefootmark{3}{Previously published in  \citetads{2013ApJ...768...77A}.
Table columns are as follows:
\begin{itemize}
\item Col. 1: The HVC name of the source following the traditional convention of galactic coordinates at the nominal
cloud center and the \vlsr\ of the cloud, for example HVC214.76+42.45+44 has $l=$214.76\dg, $b=$42.45, and \vlsr = 44 \kms.
\item Col. 2: Identification number in the Arecibo General Catalog (AGC), an internal database maintained by MH and RG. This identifier allows for easy cross--reference with the ALFALFA survey catalogs. Generally, we will use this identifier
for the UCHVCs for brevity. Footnotes in this column indicate references for previously published UCHVCs.
\item Col. 3: Equatorial coordinates of the \hi\ centroid, epoch J2000.
\item Col. 4: Recessional velocity in the heliocentric frame.
\item Col. 5: \hi\ line full width at half maximum with estimated measurement error in brackets.
\item Col. 6: Average angular diameter at the half-flux level, $\bar a$, computed as the geometric mean of the major and minor
axes of the half-power ellipse: $\sqrt{ab}$. 
\item Col. 7: Flux density integral in Jy \kms\ with measurement error in brackets
\item Col. 8: $\log$ of the \hi\ mass for an {\it assumed} distance of 1 Mpc in units of \msun
\item Col. 9: $\log$ of the representative column density, $\bar N_{HI}$,  in units of atoms cm$^{-2}$
\end{itemize}
}
}
\end{table*}

\section{Data}\label{sec:data}

\subsection{ALFALFA}
ALFALFA is an extragalactic spectral line survey 
with the ALFA receiver at the Arecibo 305m telescope.
The survey maps 7000 square degrees of sky over the
spectral range  1335 to 1435 MHz (roughly -2500 \kms\ to 17500 \kms\ for
the \hi\ line), with a spectral resolution of 25 kHz, or $\sim 5.5$ \kms\ at $z=0$
\citepads{2005AJ....130.2598G,2007AJ....133.2569G}.
The UCHVCs are identified and measured independently of the standard ALFALFA pipeline,
in order to properly account for their extended and low surface brightness nature compared
to the standard ALFALFA sources. A full reporting of this methodology is given in \citetalias{2013ApJ...768...77A}.
Here, we briefly summarize the key measured \hi\ properties from the ALFALFA \hi\ data.
After automated identification, each source is visually inspected and remeasured.
Relevant measured properties are the \hi\ centroid,
the major and minor axes of the half-flux ellipse, integrated line flux, velocity and linewidth.
The geometric mean of the half-flux ellipse is used as a representative source size.
In addition, for the UCHVCs, a representative  column density value, $\bar N_{HI}$, is calculated following:
\be
\bar N_{HI} [\mathrm{atoms\ cm}^{-2}]  =  4.4\times 10^{20} ~\bar a^{-2} ~S_{21} \,\,\,\,{\rm cm}^{-2}, \\
\ee
where $\bar{a}$ is representative source size in arcminutes and $S_{21}$ the integrated \hi\ flux density
in Jy \kms.
The final UCHVC catalog of \citetalias{2013ApJ...768...77A} is constructed by  including all sources
with $| v_{LSR}| > 120$ \kms, an \hi\ major axis
less than 30\arcmin, a signal-to-noise (S/N) greater than eight,
and that meet the isolation criteria.
The isolation criteria are one of the most critical parameters for defining the UCHVC
sample and full details are given in \citetalias{2013ApJ...768...77A}.
 The distance of a UCHVC from another source is parameterized as 
$D = \sqrt{\theta^2 + (f\delta v)^2}$,
where $\theta$ is the angular separation in degrees, $\delta v$ is the velocity separation
in \kms, and $f$ is a factor that links a velocity separation to an angular separation.
Sources must meet isolation three criteria:
(1) visual inspection shows no connection to other \hi\ structure;
(2) the UCHVCs are separated from the
classic HVCs of \citetads[][B. Wakker 2012, private communication]{1991A&A...250..509W}
 by $D>15$\dg\ for $f=$ 0.5 \dg/\kms;
(3) and the UCHVCs have 
no more than 3 neighbors  in the ALFALFA data  within $D=3$\dg\ for $f=0.2$\dg/\kms.  
In addition, a most-isolated subsample is defined where there are no more than
 3 neighbors in the ALFALFA data within  $D=10$\dg, a 
$\sim$30$\times$ larger region. 

\begin{figure*}
\centering
\includegraphics[keepaspectratio,width=\linewidth]{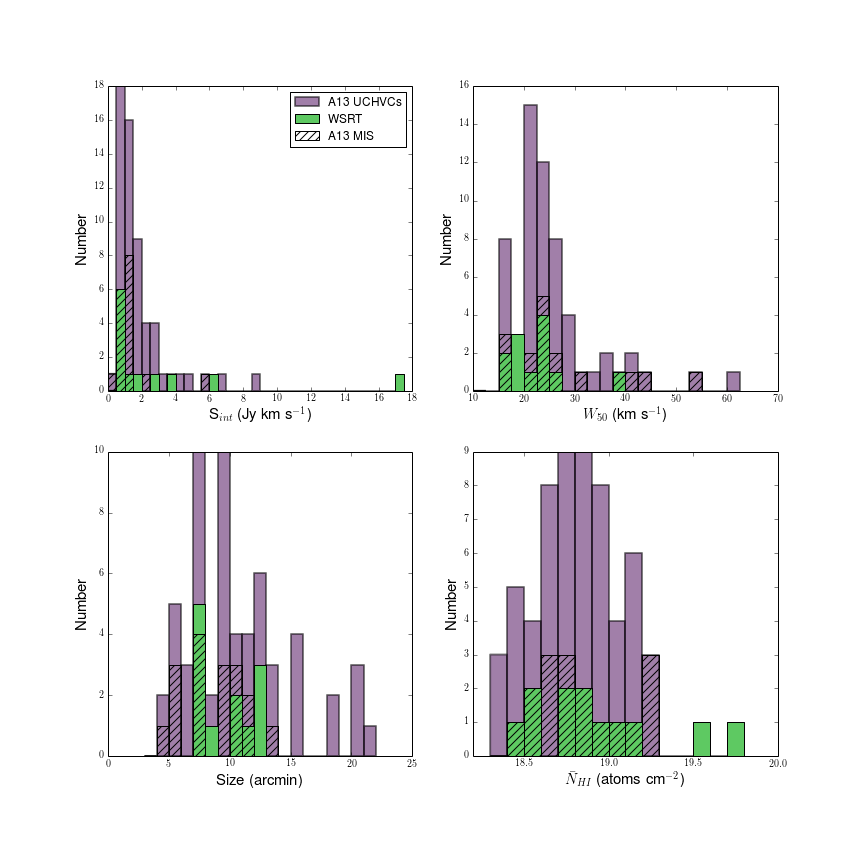}
\caption{
ALFALFA \hi\ properties of the UCHVCs presented in this work compared to the ALFALFA \hi\ properties of
the catalog of \citetalias{2013ApJ...768...77A}, including the most-isolated subsample (MIS).
}
\label{fig:histcompare}
\end{figure*}

Two of the UCHVCs, AGC\,198511 and AGC\,249393, are included in the \citetalias{2013ApJ...768...77A} catalog. A third UCHVC,
AGC\,198606, is presented independently in \citetads{2015A&A...573L...3A} as an excellent galaxy candidate.
These three UCHVCs, along with AGC\,208747, AGC\,219656, AGC\,229326, AGC\,229327
and AGC\,249525, are included in the 70\% ALFALFA catalog \citepads{2016MNRAS.457.4393J}.
The other four UCHVCs of this work have not been previously published.
The ALFALFA \hi\ properties of all of the UCHVCs of this work are reported in Table \ref{tab:hi_alfalfa}.

Figure \ref{fig:histcompare} shows the distribution of $S_{21}$, $W_{50}$, $\bar a$ and $N_{HI}$ for the
UCHVC catalog of \citetalias{2013ApJ...768...77A}, including the most-isolated subsample, and the UCHVCs observed in this work.
Generally, the UCHVCs observed here have a similar distribution of \hi\ properties to those listed in the catalog of \citetalias{2013ApJ...768...77A}.
The ALFALFA data are presented in Figures \ref{fig:hvc214.78+42.45+47}--\ref{fig:hvc28.09+71.87-142}.
The \hi\ spectra are shown in the upper left panels, and the 
upper right 
panels show total intensity \hi\ maps from ALFALFA in grayscale.

\subsection{WSRT}\label{sec:wsrtdata}
The twelve UCHVCs were observed with WSRT
over the course of two semesters under programs 13B-007 and 14A-017. 
The sources were observed using  standard 13-hour synthesis imaging tracks consisting of 
12 hours on source bracketed by half an hour on a standard calibrator.
The sources were observed for a minimum of two tracks; a few sources were 
reobserved for a third track due to poor data quality in one of the original tracks.
These observations occurred during the transition to the Apertif system and while the 
antennas were being refurbished. For the majority of observations, at least three antennas were out of
the array for testing of Apertif prototypes, and, especially for the later observations, 
more antennas were often out of the array. 
Table \ref{tab:cube_info} gives the noise value of the final data cubes for all the sources; 
the varying sensitivity of the different observations can be seen in the different noise values.
The spectral setup of the observations was a 10 MHz bandwidth divided into 2048 channels,
corresponding to a native resolution of 4.88 kHz, or $\sim$1 \kms.

\begin{table*}
\footnotesize
\caption{WSRT imaging parameters}
\label{tab:cube_info}
\centering
\begin{tabular}{llllllllll}
\hline \hline
HVC name & AGC &  \multicolumn{3}{c}{Taper} &  \multicolumn{3}{c}{Restoring beam} &  rms & $\sigma_{N_{HI}}$ \\
                   &          &  Major & Minor  & P.A. &  Major & Minor  & P.A. &$\Delta$v=4 \kms  & $\Delta$v=20 \kms \\
                   &           &   \arcsec &   \arcsec & deg  & \arcsec &   \arcsec & deg  &   mJy bm$^{-1}$ & $10^{18}$ atoms cm$^{-2}$\\
		(1) & (2) & \multicolumn{3}{c}{(3)} & \multicolumn{3}{c}{(4)} & (5) & (6)\\
\hline
HVC214.78+42.45+47 & 198606 & 265 & 195 & 115 & 209.6 & 205.7 & -43 & 1.9 & 0.4\\
                  &        & 125 & 100 & 115 & 103.1 & 102.5 & -35 & 1.6 & 1.5\\
                  &        & 60 & 60 & 90 & 62.4 & 51.5 & -166 & 1.4 & 4.3\\
HVC204.88+44.86+147 & 198511 & 240 & 185 & 105 & 204.9 & 200.8 & 76 & 2.0 & 0.5\\
                  &        & 120 & 95 & 110 & 102.5 & 100.4 & -26 & 1.5 & 1.4\\
HVC205.83+45.14+173 & 198683 & 245 & 185 & 105 & 206.9 & 204.4 & 57 & 2.1 & 0.5\\
                  &        & 120 & 95 & 110 & 101.9 & 99.8 & -31 & 1.6 & 1.6\\
HVC217.77+58.67+96 & 208747 & 255 & 185 & 110 & 208.4 & 204.9 & -36 & 2.5 & 0.6\\
                  &        & 120 & 95 & 110 & 100.0 & 99.2 & -2 & 1.7 & 1.7\\
HVC212.68+62.39+64 & 208753 & 255 & 195 & 105 & 208.8 & 205.1 & -36 & 2.5 & 0.6\\
                  &        & 120 & 90 & 100 & 100.8 & 98.7 & 59 & 1.8 & 1.8\\
HVC230.27+71.10+76 & 219663 & 260 & 190 & 110 & 208.6 & 203.6 & -43 & 3.3 & 0.8\\
                  &        & 120 & 85 & 90 & 103.6 & 101.1 & 24 & 2.7 & 2.5\\
HVC235.38+74.79+195 & 219656 & 250 & 185 & 110 & 207.1 & 202.6 & -15 & 2.3 & 0.5\\
                  &        & 120 & 95 & 110 & 100.3 & 99.3 & -4 & 1.8 & 1.8\\
HVC271.57+79.03+248 & 229326 & 250 & 190 & 110 & 206.4 & 202.7 & 55 & 2.3 & 0.5\\
                  &        & 120 & 95 & 110 & 99.7 & 98.1 & 65 & 1.9 & 1.9\\
HVC276.53+79.84+255 & 229327 & 260 & 190 & 110 & 206.1 & 201.3 & 65 & 2.4 & 0.6\\
                  &        & 120 & 90 & 110 & 102.1 & 99.6 & -19 & 1.8 & 1.7\\
HVC028.09+71.87-142 & 249393 & 255 & 205 & 110 & 206.8 & 203.2 & 66 & 3.1 & 0.7\\
                  &        & 120 & 95 & 110 & 103.1 & 102.0 & -178 & 2.4 & 2.3\\
HVC011.76+67.89+60 & 249525 & 255 & 195 & 110 & 209.1 & 205.3 & 55 & 2.3 & 0.5\\
                  &        & 120 & 95 & 110 & 101.7 & 99.5 & -15 & 1.7 & 1.7\\
                  &        & 60 & 60 & 90 & 63.2 & 53.9 & 13 & 1.4 & 4.1\\
HVC015.96+63.90+44 & 249565 & 265 & 200 & 115 & 208.5 & 203.5 & 86 & 3.5 & 0.8\\
                  &        & 120 & 90 & 110 & 100.0 & 100.0 & -50 & 2.1 & 2.1\\
                  &        & 60 & 60 & 90 & 71.7 & 53.5 & 15 & 1.8 & 4.6\\
\hline
\end{tabular}
\tablefoot{
Table columns are as follows:
\begin{itemize}
\item Col. 1: HVC name as in Table \ref{tab:hi_alfalfa}
\item Col. 2: AGC identifier as in Table \ref{tab:hi_alfalfa}
\item Col. 3: Applied taper: major axis (\arcsec), minor axis (\arcsec) and position angle (\dg)
\item Col. 4: Restoring beam: major axis (\arcsec), minor axis (\arcsec) and position angle (\dg)
\item Col. 5: Rms in a 4 \kms\ channel (mJy beam$^{-1}$)
\item Col. 6: Column density rms in units of $10^{18}$ atoms cm$^{-2}$ for an assumed linewidth of 20 \kms
\end{itemize}
}
\end{table*}

The data were reduced in Miriad following standard procedure \citepads{1995ASPC...77..433S}. 
The data were flagged manually for 
radio frequency interference (RFI),
and the bandpass and initial gain calibration used the standard calibrators. 
A continuum image of the target field was used for a phase-only self-calibration.
Line-free channels were used to 
subtract the continuum emission in the uv-plane. A first-order polynomial was used for removing the
continuum except for a few cases where a higher-order polynomial was justified.

Imaging of the data was done using the Common Astronomy Software Applications
package \citepads[CASA,][]{2007ASPC..376..127M}.
 A spectral resolution of 4 \kms\
and a spectral range of 100 \kms\ was used for all the sources.
The spectral range was centered on the source except for a few objects
close to Galactic emission in velocity space. Then, the center velocity was offset
to avoid having channels dominated
by Galactic \hi\ and to provide more signal-free channels. 
Due to the low surface brightness nature of the UCHVCs, significant tapering was required
to robustly detect emission.
All the sources were tapered to 
resolutions comparable to that of Arecibo ($\sim 210$\arcsec)
and about twice the angular resolution ($\sim$105\arcsec). 
In addition, the sources that were strongly detected at 105\arcsec\ resolution were imaged
with a 60\arcsec\ taper applied.
Table \ref{tab:cube_info} lists the applied tapers and the resulting beams for each source.

\begin{table*}
\footnotesize
\caption{WSRT \hi\ properties}
\label{tab:hi_wsrt}
\centering
\begin{tabular}{llllllllllll}
\hline \hline
HVC name & AGC & R.A.+Dec. & cz  & \w50&  $\theta_{hf}$     &\multicolumn{3}{c}{S$_{HI}$}&   \multicolumn{3}{c}{$N_{HI, peak}$}\\
         &     &           &    &       & 210\arcsec  & 210\arcsec & 105\arcsec  &60\arcsec &  210\arcsec & 105\arcsec & 60\arcsec \\
        &      &   J2000     &\kms& \kms & \arcmin    & \multicolumn{3}{c}{Jy \kms} & \multicolumn{3}{c}{ $10^{19}$ atoms cm$^{-2}$} \\ 
        (1) & (2) & (3) &(4) &(5)&(6) & \multicolumn{3}{c}{(7)} &\multicolumn{3}{c}{(8)}\\
\hline
HVC214.78+42.45+47 & 198606 & 093002.8+163813 & 51 & 24 & 11.4 & 15.3 & 14.0 & 11.8 & 4.2 & 4.9 & 6.1\\
HVC204.88+44.86+147 & 198511 & 093017.2+241119 & 151 & 15 & 7.0 & 0.86 & 0.78 & \ldots & 0.59 & 0.98 & \ldots\\
HVC205.83+45.14+173 & 198683 & 093212.8+233348 & 176 & 23 & 8.2 & 0.63 & 0.42 & \ldots& 0.46 & 1.2 & \ldots\\
HVC217.77+58.67+96 & 208747 & 103710.0+203214 & 96 & 19 & 6.8 & 1.69 & 1.57 & \ldots & 1.1 & 1.6 & \ldots\\
HVC212.68+62.39+64 & 208753 & 104931.8+235526 & 65 & 24 & 10.6 & 3.32 & 3.1 & \ldots & 0.93 & 1.4 &\ldots\\
HVC230.27+71.10+76 & 219663 & 113422.1+201212 & 73 & 13 & 5.8 & 0.65 & 0.49 & \ldots& 0.68 & 1.1 & \ldots\\
HVC235.38+74.79+195 & 219656 & 115123.7+203403 & 191 & 28 & 8.0 & 1.22 & 1.09 &\ldots& 0.67 & 1.2 &\ldots\\
HVC271.57+79.03+248 & 229326 & 122735.5+173852 & 239 & 22 & 5.5 & 0.63 & 0.58 &\ldots & 0.63 & 1.0 & \ldots\\
HVC276.53+79.84+255 & 229327 & 123223.6+175522 & 253 & 17 & 8.8 & 0.62 & 0.62 &\ldots & 0.38 & 0.88 & \ldots\\
HVC028.09+71.87-142 & 249393 &\ldots & \ldots& \ldots & \ldots &\ldots& \ldots&\ldots& 0.34\tablefootmark{a} &\ldots &\ldots\\
HVC011.76+67.89+60 & 249525 & 141752.7+173240 & 47 & 18 & 6.8 & 5.12 & 4.85 & 4.64 & 3.6 & 4.8 & 5.4\\
HVC015.96+63.90+44 & 249565 & 143556.3+170847 & 29 & 18 & 5.4 & 1.69 & 1.74 & 1.6 & 1.8 & 2.9 & 4.0\\
\hline
\end{tabular}
\tablefoot{
\tablefootmark{a}{This source is a non-detection; the peak column density value
is based on the moment zero map of the dirty data.}
\\
Table columns are as follows:
\begin{itemize}
\item Col. 1: The HVC name of the source as for Table \ref{tab:hi_alfalfa}
\item Col. 2: AGC number as in Table \ref{tab:hi_alfalfa}
\item Col. 3: Equatorial coordinates of the \hi\ centroid (epoch J2000) for the 210\arcsec\ WSRT data, derived using the task {\tt maxfit} in Miriad. 
\item Col. 4: Heliocentric recessional velocity, derived from Gaussian fitting to the WSRT 210\arcsec\ spectrum
\item Col. 5: Full width at half maximum of the WSRT 210\arcsec\ spectrum, derived from Gaussian fitting
\item Col. 6: Effective half-flux radius in arcminutes, derived as discussed in Section \ref{sec:wsrtdata}.
\item Col. 7: Integrated flux density for each angular resolution, calculated as discussed in Section \ref{sec:wsrtdata}. The uncertainty is taken to be 10\%.
\item Col. 8: Peak $N_{HI}$ for each angular resolution in units of $10^{19}$ atoms cm$^{-2}$
\end{itemize}
}
\end{table*}

Since the UCHVCs are narrow in velocity extent, show only small amounts of velocity structure, and
are generally low S/N, special care was taken in isolating the emission for cleaning.
For each spatial resolution, a single clean mask was created and used for all channels with emission. 
This mask was constructed by making a single channel image using the central velocity and velocity extent from the ALFALFA \hi\ spectrum.
This single channel image was then iteratively cleaned until a final image was reached. 
This image was then smoothed to 300\arcsec, 210\arcsec, or 120\arcsec\ resolution for the 210\arcsec, 105\arcsec, and 60\arcsec\ data, and clipped at the 2-$\sigma$ level to define the extent of source emission. The channels of the data cube with emission were determined by examining the dirty cubes and using the ALFALFA spectra as a guide.
 The single channel image mask was then used to populate these channels to create a 3-D clean mask. The data were then cleaned deeply, to half the rms value, to minimize the impact of residual flux.

Moment zero (total \hi\ intensity) maps were created over the range of channels identified
as having emission, without any pre-applied masking. 
The velocity range used for constructing
the moment zero maps is indicated in the spectra panel in Figures \ref{fig:hvc214.78+42.45+47}--\ref{fig:hvc28.09+71.87-142}.
The non-primary-beam-corrected map for the 210\arcsec\ data in contours of significance is compared
to the ALFALFA \hi\ map in Figures \ref{fig:hvc214.78+42.45+47}--\ref{fig:hvc28.09+71.87-142}. In all cases, the  ALFALFA \hi\ morphology is well matched to the
WSRT \hi\ morphology at the same resolution;
when the ALFALFA data suggest structure in the \hi\ morphology this is also seen in the WSRT data.
The primary-beam-corrected total intensity \hi\ maps in units of column density
for all imaging resolutions are also presented in Figures \ref{fig:hvc214.78+42.45+47}--\ref{fig:hvc15.96+63.90+44}. 
(AGC\,249393 is a non-detection in the WSRT data and so only a partial presentation of the WSRT data is
included in Figure \ref{fig:hvc28.09+71.87-142}.)
Peak column density values from these maps are reported in Table \ref{tab:hi_wsrt}.

The primary-beam-corrected moment zero maps were used to find the total \hi\ line flux of the sources.
Since the contribution of low level emission is important for low surface brightness objects,
the moment zero maps were smoothed 
before defining the source extent to determine the flux density.
The level of smoothing was the same as for the creation of the clean masks:
300\arcsec, 210\arcsec, and 120\arcsec\ for the 210\arcsec, 105\arcsec, and 60\arcsec\ data.
The smoothed maps were clipped at the 3-$\sigma$ level to define the source extent;
this mask can be seen in Figures \ref{fig:hvc214.78+42.45+47}--\ref{fig:hvc15.96+63.90+44}.
The emission within this region in the original resolution maps was summed to determine the 
integrated flux density.
Generally the region used for calculating the integrated line flux is much more extended than
the high significance emission.
The WSRT spectra shown in Figures \ref{fig:hvc214.78+42.45+47}--\ref{fig:hvc15.96+63.90+44} are derived using the same mask applied to
primary-beam-corrected data cubes.
Table \ref{tab:hi_wsrt} gives the final line flux values for the source for all resolutions.
Generally, the derived integrated line flux is the same at both the 210\arcsec\ and 105\arcsec\ resolution,
and reasonably close to the single-dish ALFALFA value. 
We discuss this further in Section \ref{sec:compare}.
We take the uncertainty on the flux to be 10\%; this is dominated by uncertainty
in determining the source extent.

For regions of significant emission ($> 5 \sigma$ in the 210\arcsec\ total intensity \hi\ map
and $>3\sigma$ for the higher resolution maps),
we created moment 1 and moment 2 maps, 
representing the velocity field and velocity dispersion. 
These maps are also shown in Figures \ref{fig:hvc214.78+42.45+47}--\ref{fig:hvc15.96+63.90+44},
along with  
 position-velocity slices at all resolutions derived using the CASA task {\tt impv}.
The angle of the slices was set
to highlight structure, either in \hi\ morphology or kinematics, 
and can be seen overlaid on top of the moment maps.

In addition, the effective half-flux radii of the sources were estimated by measuring the integrated flux density in increasing circular apertures until half the total WSRT \hi\ line flux is enclosed. In many cases the source structure is elongated,
and so this is a very crude measure and is reported without errors. 
However, it is still a useful metric for comparison to the ALFALFA \hi\ size. 
The WSRT half-flux apertures are reported in
Table \ref{tab:hi_wsrt} and shown on
the grayscale ALFALFA maps in Figures \ref{fig:hvc214.78+42.45+47}--\ref{fig:hvc15.96+63.90+44} 
with the ALFALFA half-power ellipse and  $\bar a$ shown for reference.


\section{Comparison of ALFALFA and WSRT derived \hi\ properties}\label{sec:compare}

\begin{table*}
\footnotesize
\caption{Comparison of ALFALFA and WSRT \hi\ properties}
\label{tab:hi_compare}
\centering
\begin{tabular}{lllllll}
\hline \hline
HVC name & AGC &$\theta_{hf}/\bar a$ &\multicolumn{2}{c}{$S_{WSRT}/S_{ALFALFA}$}&\multicolumn{2}{c}{$N_{HI,WSRT}/\bar N_{HI}$}\\
  & & & 210\arcsec &105\arcsec &210\arcsec & 105\arcsec\\
  (1) & (2) & (3) & \multicolumn{2}{c}{(4)} &\multicolumn{2}{c}{(5)}\\
\hline
HVC214.78+42.45+47 & 198606 & 0.9 & 0.9 (0.1) & 0.8 (0.1) & 0.8 (0.1) & 1.0 (0.1)\\
HVC204.88+44.86+147 & 198511 & 1.0 & 1.2 (0.1) & 1.1 (0.1) & 0.9 (0.2) & 1.5 (0.3)\\
HVC205.83+45.14+173 & 198683 & 0.8 & 0.7 (0.1) & 0.5 (0.1) & 1.3 (0.3) & 3.3 (0.8)\\
HVC217.77+58.67+96 & 208747 & 0.6 & 0.6 (0.1) & 0.6 (0.1) & 1.1 (0.2) & 1.6 (0.4)\\
HVC212.68+62.39+64 & 208753 & 0.8 & 0.8 (0.1) & 0.8 (0.1) & 0.9 (0.1) & 1.3 (0.3)\\
HVC230.27+71.10+76 & 219663 & 0.8 & 0.9 (0.1) & 0.7 (0.1) & 1.1 (0.2) & 1.7 (0.5)\\
HVC235.38+74.79+195 & 219656 & 1.1 & 1.4 (0.2) & 1.2 (0.1) & 1.0 (0.2) & 1.7 (0.4)\\
HVC271.57+79.03+248 & 229326 & 0.7 & 0.8 (0.1) & 0.8 (0.1) & 1.1 (0.2) & 1.8 (0.5)\\
HVC276.53+79.84+255 & 229327 & 0.8 & 0.7 (0.1) & 0.7 (0.1) & 1.2 (0.3) & 2.7 (0.7)\\
HVC028.09+71.87-142 & 249393 &\ldots & \ldots&\ldots& 1.2\tablefootmark{a} &\ldots\\
HVC011.76+67.89+60 & 249525 & 0.8 & 0.8 (0.1) & 0.8 (0.1) & 0.9 (0.1) & 1.2 (0.2)\\
HVC015.96+63.90+44 & 249565 & 0.7 & 1.0 (0.1) & 1.0 (0.1) & 1.3 (0.2) & 2.1 (0.4)\\
\hline
\end{tabular}
\tablefoot{
\tablefootmark{a}{This source is a non-detection; the peak column density value
is based on the moment zero map of the dirty data.}
\\
Table columns are as follows:
\begin{itemize}
\item Col. 1: The HVC name of the source as for Table \ref{tab:hi_alfalfa} 
\item Col. 2: AGC number as in Table \ref{tab:hi_alfalfa}
\item Col. 3: The ratio of the  WSRT half-flux size to the ALFALFA half-flux size
\item Col. 4: The ratio of the WSRT integrated flux density to the ALFALFA integrated flux density
for both the 210\arcsec and 105\arcsec data.
\item Col. 5: The ratio of the peak WSRT column (for both the 210\arcsec\ and 105\arcsec\ data)
to the ALFALFA $\bar N_{HI}$.
\end{itemize}
}
\end{table*}

\begin{figure*}
\centering
\includegraphics[keepaspectratio,width=\linewidth]{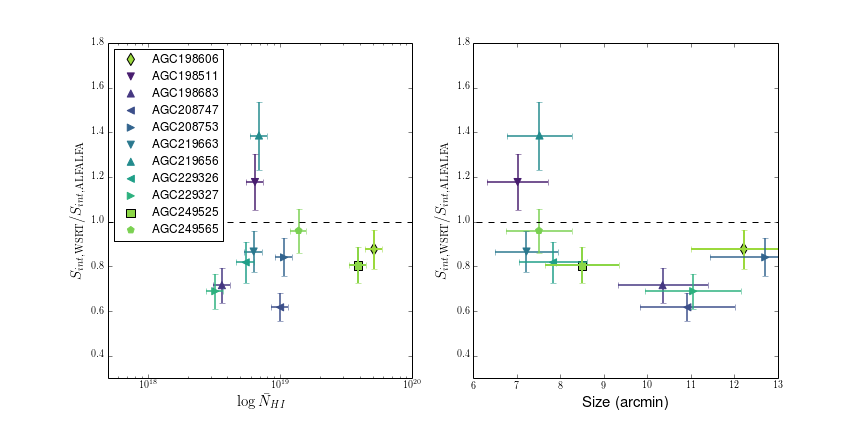}
\caption{Fraction of ALFALFA flux recovered  in the 210\arcsec\ WSRT observations as a function of $\bar N_{HI}$ and size.}
\label{fig:fluxrecov}
\end{figure*}

Table \ref{tab:hi_compare} lists the comparison of WSRT \hi\ properties to the ALFALFA \hi\
properties. 
The first thing to note is that the flux recovery is generally excellent (more than 50\% in all cases)
and there is generally very little change between the 210\arcsec\ and 105\arcsec\ WSRT data.
This is due in large part to the expansive masks as a result of smoothing used for defining source extent.
If the sources had been clipped directly on a threshold, the flux recovery would be much less, especially for the 105\arcsec\ data
as much of the flux is in low surface brightness emission that is on the level of the noise.
Figures \ref{fig:hvc214.78+42.45+47}--\ref{fig:hvc15.96+63.90+44} illustrate this; the masks that define source extent are much broader than where the significant emission
is located in most cases.
Traditional wisdom says that interferometers miss flux because of the lack of short spacings but in many cases it may be an issue of lack
of sensitivity to (or properly including) low column density emission rather than extended emission that is resolved out.
Figure \ref{fig:fluxrecov} shows the fraction of line flux recovered in the WSRT 210\arcsec\ data relative
to the ALFALFA data as a function of $\bar N_{HI}$ and
size of the source. Generally, the sources with the highest peak column densities and the smallest sizes have the best flux recovery.

\begin{figure*}
\centering
\includegraphics[keepaspectratio,width=\linewidth]{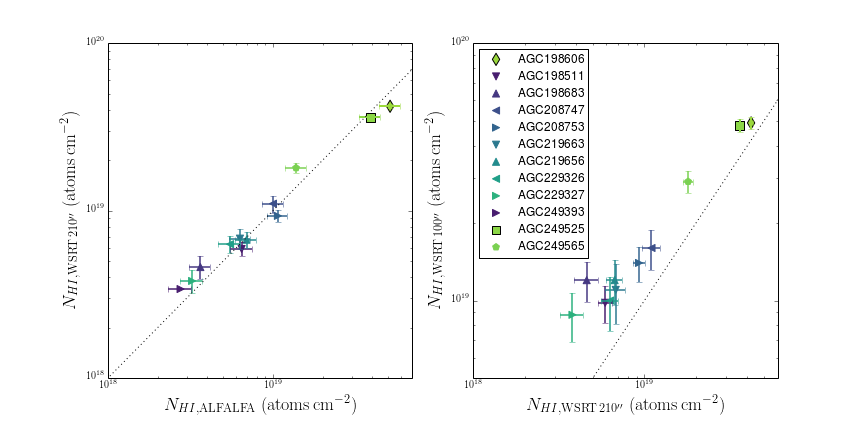}
\caption{Left panel: ALFALFA $\bar N_{HI}$ versus the peak $N_{HI}$ from the 210\arcsec\ WSRT data. Right panel:  $N_{HI}$
at 105\arcsec\ resolution compared to 210\arcsec.}
\label{fig:nhi}
\end{figure*}

 The left panel of Figure \ref{fig:nhi} shows that the ALFALFA $\bar N_{HI}$ value agrees extremely well
 with the WSRT 210\arcsec\ peak $N_{HI}$,
  motivating the use of this average ALFALFA value.
The right panel of Figure \ref{fig:nhi} also shows the peak $N_{HI}$ from the 105\arcsec\ WSRT data compared to that from the 210\arcsec\
 data. All the sources have higher peak $N_{HI}$ at higher angular resolution as expected.

 Table \ref{tab:hi_compare} also lists the comparison of the effective half-flux size derived from the WSRT data to that from the ALFALFA data.
While these sizes are in rough agreement, the sizes derived from the WSRT data tend to be smaller. 
That is consistent with the fact that WSRT data typically have
 smaller line flux values  and the half-flux size is derived self-consistently for the WSRT data.
Figures \ref{fig:hvc214.78+42.45+47}--\ref{fig:hvc28.09+71.87-142} also show that the \hi\ morphology seen in the ALFALFA maps is matched by the morphology in the WSRT \hi\ maps. 
 When the ALFALFA data show elongated structure or evidence for multiple clumps, that structure is also seen in the WSRT \hi\ data at
 the Arecibo resolution. Unfortunately, the reported parameters of the ALFALFA half-power ellipse, 
 especially the position angle, do not always match the
 structure. This is consistent with what was seen in  \citetads{
2014PhDT........16A} where by measuring artificial sources it was shown that the average size of an UCHVC, $\bar a$, was an accurate measurement but that the measured ellipticity did not necessarily match the true ellipticity.

\section{Discussion}

\subsection{The nature of the UCHVCs}\label{sec:nature}

\begin{figure*}
\centering
\includegraphics[keepaspectratio,width=\linewidth]{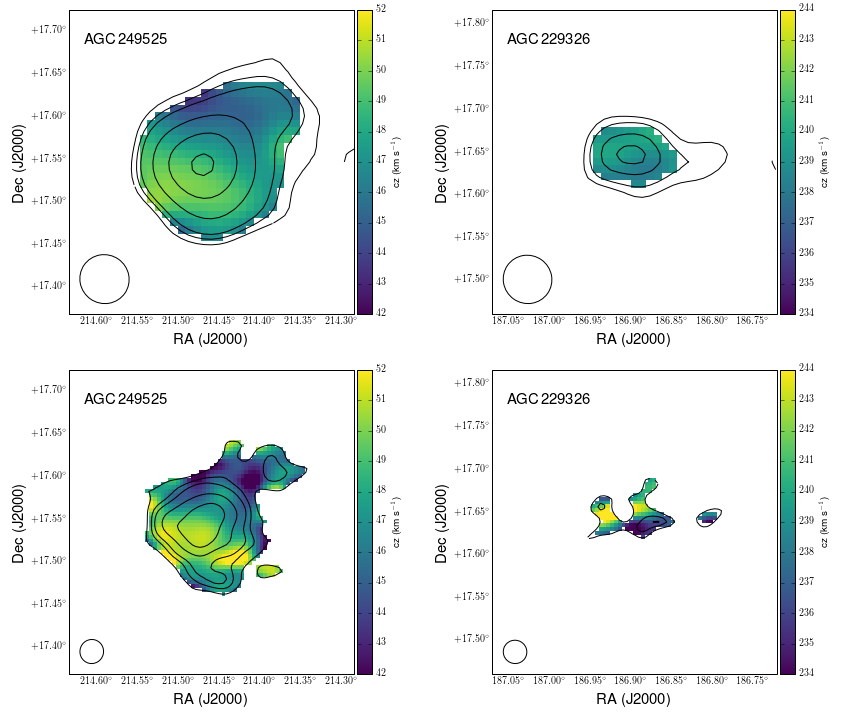}
\caption{Velocity fields with \hi\ column density contours of AGC\,249525 (left) and AGC\,229326 (right)
at 210\arcsec\ (top) and 105\arcsec\ (bottom) resolution. 
AGC\,249525 is representative of the potential galaxy candidates and
AGC\,229326 of the likely Galactic halo clouds.
The column density contours for AGC\,249525
for the 210\arcsec\ and 105\arcsec\ data are [4, 6, 8, 15, 25, 35] and [9, 15, 20, 30, 40] $\times
10^{18}$ atoms cm$^{-2}$. For AGC\,229326 the column density levels are
[2.5, 3.5, 5, 6] and [7, 9, 10] $\times 10^{18}$ atoms cm$^{-2}$.}
\label{fig:examplesources}
 \end{figure*}

 The WSRT observations reveal that the UCHVCs fall into two categories:
 likely Galactic halo clouds and potential galaxy candidates.
  These two source populations are distinguished from each other based on a 
  combination of \hi\ morphology and kinematics.
 All the sources show a smooth \hi\ morphology in the 210\arcsec\ resolution data;
 this is consistent with their selection as UCHVCs from the ALFALFA data.
 The first set of sources, likely Galactic halo clouds,
 are distinguished by 
 a lack of ordered velocity motion at all spatial resolutions and 
 breaking into clumps of low
 column density \hi\ contained in a much more diffuse envelope in the 105\arcsec\
 resolution data.
 This is consistent with previous observations of (compact) HVCs,
and the likely explanation for these sources is that they are
clouds of \hi\ in the halo of the MW at typical
distances of a 100 kpc
\citepads[e.g.,][]{2001A&A...370L..26B,2002A&A...391...67D,2005A&A...432..937W}.
 In contrast, the potential galaxy candidates have 
 ordered velocity motion at all angular resolutions and
 a smooth \hi\ morphology in the 105\arcsec\
 resolution data. 
 The potential galaxy candidates also have higher column density values,
 allowing them to be imaged at 60\arcsec\ resolution, 
 where they maintain the smooth \hi\ morphology and ordered velocity motion.
 Figure \ref{fig:examplesources} illustrates the comparison between the 210\arcsec\
 and 105\arcsec\ resolution data for an example likely Galactic halo cloud
 and one of the potential galaxy candidates.

\begin{figure*}
\centering
\includegraphics[keepaspectratio,width=\linewidth,clip=true,trim=2cm 2cm 2cm 2cm]{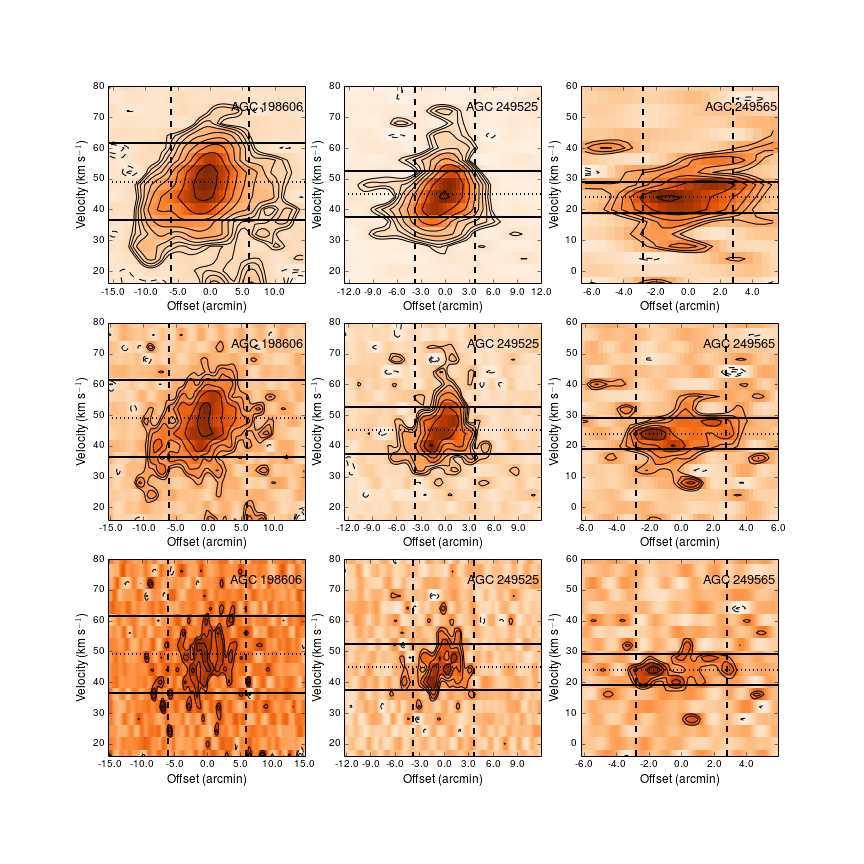}
\caption{Position-velocity slices along the direction of ordered motion for the three
galaxy candidates at 210\arcsec, 105\arcsec\, and 60\arcsec\ resolution (top to bottom).
From left to right the sources are AGC\,198606, AGC\,249525, and AGC\,249565.
The contours start at 2 $\times$ rms of the data cube
 and increase by $\sqrt{2}$; negative contours with the same spacing are also shown.
 The solid horizontal black lines indicate the extent of velocity motion,
 and the dashed vertical lines indicate the spatial extent over which it occurs.}
 \label{fig:pvslices}
\end{figure*}

 The \hi\ column density, velocity field, and velocity dispersion maps and the
 position-velocity slices used to determine the nature of the UCHVCs are 
presented in Appendix \ref{sec:sources}.
We determined that three sources 
are potential galaxy candidates. 
The remainder of the UCHVCs are considered likely 
Galactic halo clouds.
The two sources
from this work included in \citetads{2015A&A...575A.126B} and \citetads{2015ApJ...806...95S}, 
AGC\,198511 and AGC\,249393, show  the morphology and lack of ordered velocity motion
that is typical of Galactic halo clouds.
This is consistent with the fact that these sources have no detected optical counterparts,
down to strict limits \citepads{2016A&A...591A..56B}.
 AGC\,198606 (also presented as a galaxy candidate in \citetads{2015A&A...573L...3A})
and AGC\,249525 are excellent galaxy candidates with a smooth
\hi\ morphology and evidence for ordered velocity motion at all
spatial resolutions considered.
 AGC\,249565 is a possible galaxy candidate,
 but 
  the evidence for its ordered velocity motion is weaker,
  especially at higher angular resolution,
 and  it may break into \hi\ clumps in a more
 diffuse envelope in the 60\arcsec\ resolution data,
  although this could be a S/N limitation.
  The ordered velocity motion is key for classifying a UCHVC as a galaxy
  candidate, and in
  Figure \ref{fig:pvslices} we show the position-velocity slices for all 
 three galaxy candidates to illustrate their ordered velocity motion.

\subsection{Identifying galaxy candidates}

The goal of this work is to identify \hi\ clouds that are good candidates to represent
(nearly) starless gas in dark matter halos. 
In Section \ref{sec:nature}, we determined that AGC\,198606 and AGC\,249525 were
excellent candidates to represent gas in dark matter halos while AGC\,249565 was
a potential third candidate.
The majority of the UCHVCs (9/12) 
 lack any kinematic structure and
show
a \hi\ morphology at higher angular resolutions that is consistent with local Galactic halo clouds.
This is not surprising as there are many potential formation mechanisms for HVCs and the majority of
them are Galactic processes. An important step is to identify the \hi\ properties
by which the best galaxy candidates may be recognized.
The targets were selected for WSRT observations on the basis of four criteria:
isolation, compact size, large recessional velocity, or large average column density.

The three potential galaxy candidates were not selected for observations based on
their size or isolation.
While they are  relatively compact
and isolated, they are not distinguished from the rest of the UCHVC population by these two criteria.
Instead,
they are most clearly distinguished by having high column densities,
as can be seen in Figure \ref{fig:nhi}.
In fact,
the two best candidates, AGC\,198606 and AGC\,249525,
 have column densities higher than any source included in \citetalias{2013ApJ...768...77A}. 
While the peak column densities for these sources are higher than the other UCHVCs
at all resolutions, their peak column density increases by less from the 210\arcsec\ to
105\arcsec\ data (right panel of Figure \ref{fig:nhi}). 
This is consistent with their morphology in that the UCHVCs with the largest
changes in peak column densities are those that show the most clumpiness at
higher angular resolution.

\begin{figure*}
\centering
\includegraphics[keepaspectratio,width=\linewidth]{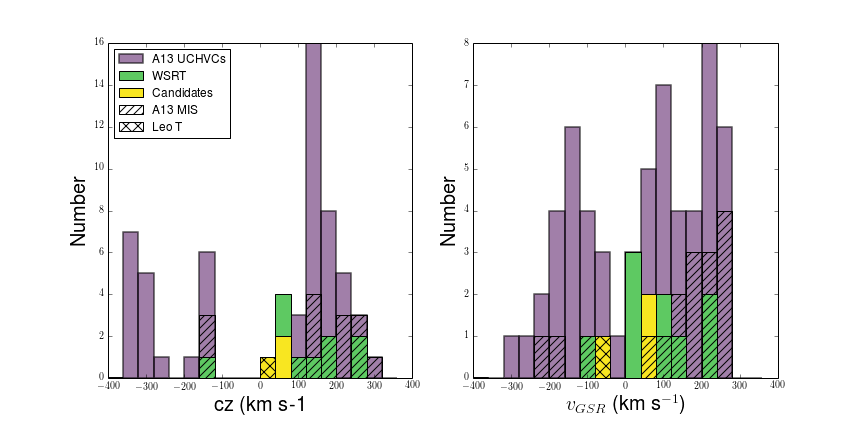}
\caption{ Distribution of velocities in heliocentric and Galactic standard of rest
frames for the \citetalias{2013ApJ...768...77A} UCHVCs and the UCHVCs of this work, with the good galaxy candidates
highlighted in yellow.}
\label{fig:histcandsvel}
\end{figure*}

Figure \ref{fig:histcandsvel} shows the recessional velocities for all the candidates
and the \citetalias{2013ApJ...768...77A} sample in two different frames: helicoentric ($cz$), 
and Galactic standard of rest ($v_{GSR}$). 
Contrary to our expectations when selecting targets,
the highest recessional velocity targets are not the best candidate galaxy candidates.
Instead, the best candidates are those at low recessional velocity, with
the three best candidates having $|v_{LSR}| < 120$ \kms,
outside the original selection selection criteria of \citetalias{2013ApJ...768...77A}.
It is worth noting that Leo T has a low recessional velocity ($cz=35$ \kms),
and we indicate its position in Figure \ref{fig:histcandsvel}. 
The picture that 
the best candidate galaxies are at low
recessional velocities is consistent with  theoretical work by \citetads{2014MNRAS.438.2578G}.
They find that never-accreted halos (the most likely to be overlooked gas-rich dwarf galaxies)
in Local Group analogs are most likely to have radial velocities relative to their host galaxy
with an amplitude $<$ 150 \kms\ (e.g., $|v_{GSR}|<150$ \kms).
This is in contrast to the work of \citetads{2015ApJ...808..136D}
who find that \hi\ clouds distinguished as velocity outliers are
most likely to have a GALEX counterpart.
However these are systems with apparent stellar counterparts 
and likely lie at larger distances within the Local Volume, rather than nearby in the Local Group.

Overall, the 
clearest distinguishing feature of the best galaxy candidates is their
high column density (albeit still low for galaxies).
They also tend to be small, isolated and at low velocity, but none of those criteria
are sufficient to identify the best candidates.
The fact that the best candidates are at low recessional velocities means that 
identifying these sources will be strongly complicated by the presence of the Galactic
\hi\ foreground.

\subsection{Properties of galaxy candidates}\label{sec:gal}

In this section, we assume AGC\,198606, AGC\,249525, and AGC\,249565 
do in fact represent gas in dark matter halos and examine what their properties would be.

The strongest evidence that these three systems potentially represent gas
in dark matter halos is the ordered velocity motion, which can be interpreted
as rotation of the gas. The extent of this velocity motion is indicated in the position-velocity slices
in Figure \ref{fig:pvslices} and is in total 25, 15, and 10 \kms\ for AGC\,198606, AGC\,259525, and
AGC\,249565.
The magnitude of the velocity motion was determined to represent the global bulk velocity of the gas;
gas exists at velocities beyond the extents indicated in Figure \ref{fig:pvslices} 
due to the dispersion of the gas about the global bulk velocity.

In order to interpret the bulk velocity motion as a rotation velocity, 
the inclination of the system must be taken into account. 
For each candidate, a column density level at which the velocity motion could be reliably traced in 
both the 210\arcsec\ and 105\arcsec\ data was empirically determined.
This physical extent is shown in Figure \ref{fig:pvslices} by the vertical dashed lines.
Due to lower sensitivity to low column density emission, the 60\arcsec\ data typically does
not trace the velocity motion to the same spatial extent as the 210\arcsec\ and 105\arcsec\ data.
The axial ratio of each system was found by fitting an ellipse at this spatial extent
to both the 105\arcsec\ and 210\arcsec\ data.
For converting to an inclination,
these systems are assumed to be thick disks
 with an intrinsic axial ratio of 0.6
\citepads{2010MNRAS.404L..60R}; changing the assumed intrinsic axial ratio
by 10\% changes the derived inclination angles by a similar amount.
For AGC\,198606,
 the velocity motion is reliably traced to the $2\times10^{19}$ atoms cm$^{-2}$ level.
The fitted ellipse has a major axis of 12\arcmin\ $\pm$ 1\arcmin\ and  the axial ratio is 0.75, which corresponds to an inclination of 56\dg.
The corresponding rotation velocity is $15^{+4}_{-1}$ \kms, where the errors
correspond to a 10\% uncertainty on the axial ratio.
For AGC\,249525  the $1.5 \times 10^{19}$ atoms cm$^{-2}$ level was used,
which has a major axis of 7.5\arcmin\ $\pm$ 0.1\arcmin\ and an axial ratio of 0.91,
corresponding to an inclination of 30\dg.
Then the rotation velocity is 15$_{-2}^{+6}$ \kms.
And for AGC\,249565 the extent of the velocity motion was seen to the $1 \times10^{19}$ atoms
cm$^{-2}$ level.
This source shows the very unusual behavior that it has a larger major axis
at this level in the 105\arcsec\ data than the 210\arcsec\ data. We take the
major axis of 5.6\arcmin\ $\pm$ 0.1\arcmin\ from the 210\arcsec\ data as the extent for this source.
The axial ratio is 0.86 and the rotation velocity is 8$^{+4}_{-2}$ \kms.
The accuracy of these values is limited by the determination of the velocity extent,
assumption that the velocity gradient 
represents rotation, and the assumed inclination of the system.
This is particularly problematic for AGC\,249525 as its axial ratio is
close to unity and small changes in the fitted axial ratio greatly impact the
final rotation velocity.
These velocity values are comparable to those seen in other low mass dwarf galaxies,
such as Leo P, Pisces A and B, and the SHIELD galaxies
 \citepads{2014AJ....148...35B,2016A&A...587L...3C,McNichols}.

These rotation velocities can be used to constrain the dynamical mass of the
systems.
These systems have low rotation
velocities, on the order of their velocity dispersion.
Thus, we follow \citetads{1996ApJS..105..269H} and explicitly include
the dynamical support of velocity dispersion of the gas when
calculating the dynamical mass within a given radius:
\be
M_{dyn} =2.325 \times 10^5 \,M_{\odot} \,\frac{V_{rot}^2 + 3\sigma_z^2}{\mathrm{km^2 \,s^{-2}}}
\, \Bigg(\frac{r}{\mathrm{kpc}}\Bigg).
\ee
We take representative velocity dispersion values from the moment two maps.
For AGC\,198606 this is 9 \kms, and for AGC\,249525 and AGC\,249565 this is
7 \kms.

In order to calculate the dynamical mass or understand
any of the other intrinsic properties of these systems,
a distance is necessary.
The most straight-forward way to  obtain a distance 
is to detect a stellar counterpart.
All three of these systems lie out side the $\alpha.40$ footprint and so are not
considered in the works of \citetads{2015A&A...575A.126B} and \citetads{2015ApJ...806...95S}.
In \citetads{2015A&A...573L...3A} we argued that due
to its small separation in position and velocity space from Leo T
that AGC\,198606 is likely located at a similar distance.
In subsequent work, \citetads{2015ApJ...811...35J} found a tentative stellar counterpart (92\% confidence)
at a distance of 383 kpc, consistent with AGC\,189606 being physically associated with Leo T.
Using the slightly higher \hi\ integrated flux density value of this work,
the \hi\ mass at a distance of 383 kpc is $5.3 \times 10^5$ \msun,
and the system
is extremely \hi\ dominated, with  \mhi/\mstar $>50$.

We can use a similar philosophy to try and constrain the distances to AGC\,249525 and AGC\,249565.
These two sources are located 4.3\dg\ from each other so we consider their potential neighbors together.
Within 10\dg\ and 200 \kms\ there are three galaxies: Bootes I at distance of 66 kpc \citepads{2006ApJ...653L.109D},
Bootes II at a distance of 46 kpc \citepads{2009ApJ...690..453K}, and UGC\,9128 at a distance of 2.27 Mpc
\citepads{2013AJ....146...86T}.
Given the observed morphological segregation in the Local Group \citepads{2014ApJ...795L...5S},
these systems are  unlikely to be associated with Bootes I and II, but could be associated with UGC\,9128.
Thus we can adopt 2 Mpc as a  representative upper distance for these two sources.
As a representative lower distance, we take 0.4 Mpc, the
distance of Leo T and AGC\,198606.

\begin{table*}
\footnotesize
\caption{Properties of galaxy candidates}
\label{tab:galprops}
\centering
\begin{tabular}{lllllll}
\hline \hline
Name & Distance & \mhi & $v_{rot}$ & $r_{HI}$ & \mdyn & \mhi/\mdyn  \\
      &  Mpc & $10^5$ \msun & \kms & kpc &$10^7$ \msun \\
      (1) &(2) &(3) &(4) &(5) &(6) &(7)\\
\hline
AGC\,198606\tablefootmark{a} & 0.383 & $5.3$ & 15$^{+4}_{-1}$ & 0.66  & $7$& 0.008 \\
AGC\,249525\tablefootmark{b} &  $0.4-2$ 
                      & $1.9-48$ 
                      & 15$^{+6}_{-2}$  
                      & 0.44  --2.2 
                      & $4-20$
                      &$0.005-0.02$ \\
 AGC\,249565\tablefootmark{c} & $0.4-2$
 			& $0.64-16$
			& 8$^{+4}_{-2}$
			& 0.33  -- 1.6
			&$ 2-8$ 
			&$0.003-0.02$ \\
 \hline
Leo T\tablefootmark{d} & 0.42 
          &2.8
          &-- 
          &0.3 
          & 0.33
          &0.085 \\
Leo P\tablefootmark{e} &  1.62
         & 8.1
         & 15
         & 0.5
         & 2.5
         & 0.032\\ 
\hline
\end{tabular}
\tablefoot{
\tablefoottext{a}{\hi\ extent measured at the $2 \times 10^{19}$ atoms cm$^{-2}$ level.}
\tablefoottext{b}{\hi\ extent measured at the $1.5 \times 10^{19}$ atoms cm$^{-2}$ level.}
\tablefoottext{c}{\hi\ extent measured at the $1 \times 10^{19}$ atoms cm$^{-2}$ level.}
\tablefoottext{d}{Values from \citetads{2008MNRAS.384..535R}.}
\tablefoottext{e}{Values from \citetads{2014AJ....148...35B} and \citetads{2015ApJ...812..158M}.}
Table columns are as follows:
\begin{itemize}
\item Col. 1: Name of the system
\item Col. 2: Known distance, or range of plausible distances, in Mpc
\item Col. 3: \hi\ mass in units of $10^5$ \msun
\item Col. 4: Rotation velocity in \kms
\item Col. 5: \hi\ radius in kpc. For the UCHVCs this is the extent to which the velocity motion
can be reliably traced in the 210\arcsec\ and 105\arcsec\ data. 
\item Col. 6: Dynamical mass in units of $10^7$ \msun, derived as described in the text above.
\item Col. 7: The ratio of \hi\ mass to dynamical mass.
\end{itemize}
}
\end{table*}

Table \ref{tab:galprops} summarizes the properties of these sources for
the relevant distances, with Leo T and Leo P given for reference.
AGC\,198606 is physically bigger than Leo T and Leo P (albeit measured at a much
lower column density level) while its \hi\ mass is intermediate between the two galaxies.
Its rotational velocity is similar to that of Leo P but its dynamical mass is larger due
to its larger physical size. 
AGC\,249525
has an
\hi\ mass slightly smaller than Leo T and an \hi\ size between that of Leo T and Leo P
at the lower bound of the distance range considered for it.
Its rotational velocity is comparable to Leo P and it has a similar dynamical mass.
At the upper end of the distance range considered, AGC\,249525 is much larger than Leo T
and Leo P in terms of \hi\ mass, \hi\ size, and dynamical mass.
At its closest plausible distance, AGC\,249565 has a quarter of the \hi\ mass of Leo T
but a similar \hi\ size.
At its furthest considered distance, AGC\,249525 has twice the \hi\ mass of Leo P and
three times the \hi\ extent. 

These three candidate galaxies are distinguished from Leo T, Leo P and other
low mass galaxies by the extremes
of their baryonic component: they have a minimal stellar component, are extremely dark matter
dominated, and have extremely low peak column densities.
AGC\,198606 has a tentative stellar component with a mass of only  $\sim 10^4$ \msun\ \citepads{2015ApJ...811...35J};
AGC\,249525 and AGC\,249565 have no known stellar counterpart.
For all distances considered for AGC\,198606, AGC\,249525 and AGC\,249565, they
are extremely dark matter dominated objects, with \mhi/\mdyn $\le$ 0.02 in all cases.
This is lower than the \mhi/\mdyn\ value of Leo P, where the stellar population also contributes 
significantly to the total baryon mass ($M_{bary}$/\mdyn $=$ 0.05). 
The peak column densities of these three candidates with a 60\arcsec\ beam (0.1 kpc at 400 kpc
or 0.6 kpc at 2 Mpc)
are $4-6 \times 10^{19}$ atoms cm$^{-2}$. 
The least resolved SHIELD galaxies have a physical resolution of $\sim 0.6$ kpc
but their peak column densities are almost an order of magnitude higher \citep{Teich}.

\begin{figure*}
\centering
\includegraphics[width=\linewidth,keepaspectratio]{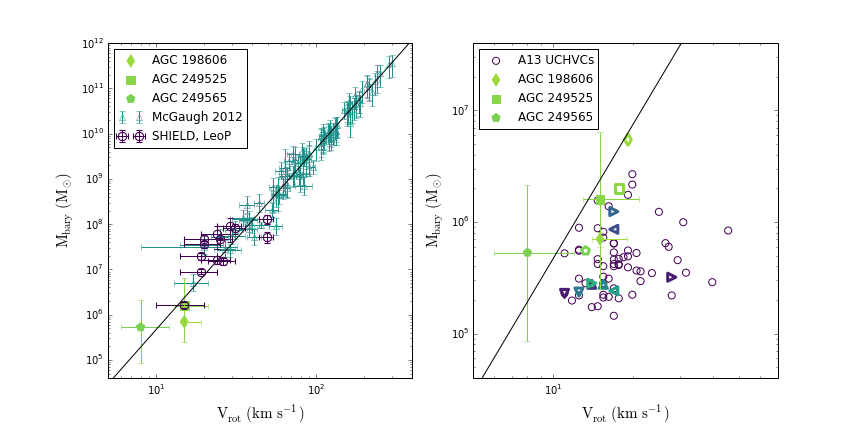}
\caption{Left: AGC\,198606, AGC\,249525 and AGC\,249565 (the three galaxy candidates) shown on the Baryonic Tully-Fisher relation (BTFR) of \citetads{2012AJ....143...40M}  with Leo P and the SHIELD galaxies \citepads{2014AJ....148...35B, McNichols} shown for extension to low rotational velocities.
Right:  UCHVCs of \citetalias{2013ApJ...768...77A} (open circles) and this work (open symbols, same colors and
shapes as in Figures \ref{fig:fluxrecov} and \ref{fig:nhi}) shown on the BTFR based on their single-dish ALFALFA properties.
The galaxy candidates are shown based on their rotation velocities derived in this work.}
\label{fig:btf}
\end{figure*}

The left panel of Figure \ref{fig:btf} places these three candidate galaxies in the context of the
baryonic Tully-Fisher relation (BTFR), a tight observed correlation
between the baryonic mass of a galaxy and its maximum rotational velocity.
We show the relation of 
\citetads{2012AJ....143...40M} along with the galaxies used in that work. We also include the
SHIELD galaxies and Leo P for an extension to lower masses 
\citepads{2014AJ....148...35B, McNichols}.
The three candidate galaxies of this work are placed on this relation using
 the rotation velocities
derived in this work and with baryonic masses
consisting of only a neutral gas component
(the \hi\ mass multiplied by a 1.33 correction factor to account for Helium). 
For AGC\,198606 we use the distance from \citetads{2015ApJ...811...35J}
of $383\pm10$ kpc and include the uncertainty in the neutral gas mass of 10\% (based
on the uncertainty of the WSRT integrated flux) in the vertical error bars;
the potential stellar component is neglected as it negligible compared to the gas mass.
For AGC\,249525 and AGC\,249565, we use a representative distance of 1 Mpc and the
vertical errors bars indicate the range of distances considered, $0.4-2$ Mpc.
These three candidate galaxies are consistent with extending the BTFR to lower
rotation velocities. AGC\,198606 and AGC\,249525 occupy the same region as Leo P
and AGC\,249565 extends the relation to even lower rotation velocities.

In order to test the significance of the three candidate galaxies lying on the BTFR, in
the right panel of Figure \ref{fig:btf} we place all the UCHVCs on the BTFR based
on their single-dish ALFALFA \hi\ properties.
The baryonic mass is the \hi\ mass for an assumed distance
of 1 Mpc multiplied by a correction factor of 1.33 to account
for Helium, and $v_{rot}$ is approximated as $\sqrt{3}\sigma$, where $\sigma$ is the 
line-of-sight
velocity dispersion of the gas based on the $W_{50}$ value.
The general UCHVC population occupies a region of parameter
space below the BTFR.
Interestingly, the candidate galaxies are among the sources that scatter closest
the BTFR based on single-dish properties.

\subsection{Implications for future surveys}
In line with previous work, these observations highlight the power of resolved \hi\ observations for addressing 
questions as to the nature of (ultra-) compact HVCs and whether they can represent gas in
dark matter halos 
\citepads[e.g.,][]{2002A&A...391...67D,2004A&A...426L...9B,2005A&A...432..937W,2005A&A...436..101W}.
Of the twelve clouds identified in the single-dish ALFALFA survey, the higher angular resolution WSRT 
observations show that only three of the objects are potential candidates to represent gas in dark matter halos.
With future large-field surveys planned with interferometers (e.g., Apertif and ASKAP),
many more of these objects will be detected and immediately distinguished as galaxy candidates.
If AGC\,198606, AGC\,249525, and AGC\,249565 do indeed represent what we might expect for gas-rich
(nearly) starless galaxies in the Local Universe, we can use them as guides for how to find more of these objects
in future surveys.
Importantly, while they are the highest column density objects studied here, their peak column densities are
lower  than that typically found in low mass galaxies
by an order of magnitude.
In order to robustly detect and image these objects in future surveys, a special handling of the data
with strong angular smoothing will be called for.

\section{Summary}

We present WSRT \hi\ observations of twelve UCHVCs identified in the ALFALFA \hi\ survey
as good candidates to be gas-bearing dark matter halos. 
Our key results are as follows:
\begin{itemize}
\item The flux recovery from the WSRT observations compared to the ALFALFA data is generally excellent.
This is due to the use of a smoothed image to define the source extent to ensure that low column density emission
on the level of the noise is included in the calculation of the integrated line flux.
\item Two of the twelve UCHVCs, AGC\,198606 and AGC\,249525, are excellent candidates to represent gas in dark matter halos. 
A third UCHVC, AGC\,249565, is  a possible candidate.
The property that most distinguishes the galaxy candidates 
 from the rest of the UCHVC population is having a higher average column density in the
ALFALFA \hi\ data.
\item In addition, the best galaxy candidates have low recessional velocities. This is consistent with theoretical
predictions for the Local Group, but means identifying the best candidates will be challenging as they must
be distinguished from the Galactic \hi\ foreground.

\item AGC\,198606 is an excellent galaxy candidate, as discussed previously in \citetads{2015A&A...573L...3A}
and \citetads{2015ApJ...811...35J}.
Based on the \hi\ properties derived in this work, and the distance of \citetads{2015ApJ...811...35J},
we find it has an \hi\ mass of $5.3 \times 10^5$ \msun, a rotational velocity of $15^{+4}_{-1}$ \kms,
a radius of 0.66 kpc, and a dynamical mass of $7 \times 10^7$ \msun.

\item AGC\,249525 is the other best galaxy candidate. 
It has a rotation velocity of $15^{+6}_{-2}$ \kms, and for the plausible range of distances $0.4-2$ Mpc,
it has an \hi\ mass of $1.9 - 48 \times10^5$ \msun, an \hi\ radius of $0.44-2.2$ kpc, 
and a dynamical mass of $4-20 \times 10^7$ \msun.

\item AGC\,249565 is the third potential galaxy candidate, although we emphasize it is much less likely
to represent gas in a dark matter halo than the other two. 
This source has a rotation velocity of $8^{+4}_{-2}$.
For the assumed distance range of $0.4-2$ Mpc, 
it has  an \hi\ mass of $0.64-16 \times10^5$ \msun, an \hi\ radius of $0.33-1.6$ kpc,
and a dynamical mass of $2-8\times10^7$\msun.
\item Future \hi\ surveys with interferometers will be able to distinguish galaxy candidates immediately on the basis
of \hi\ morphology and kinematics but must make sure that the data handling allows for for sensitivity to low
column density emission.

\end{itemize}


\begin{acknowledgements}
We thank the anonymous referee for valuable input.
The Westerbork Synthesis Radio Telescope is operated by the ASTRON (Netherlands Institute for Radio Astronomy) with support from the Netherlands Foundation for Scientific Research (NWO).
The Arecibo Observatory is operated by SRI International under a cooperative agreement with the National Science Foundation (AST-1100968), and in alliance with Ana G. M\'{e}ndez-Universidad Metropolitana, and the Universities Space Research Association.
EAKA is supported by TOP1EW.14.105, which is financed by the Netherlands Organisation for
Scientific Research (NWO).
The ALFALFA work at Cornell is supported by NSF grants AST-0607007 and AST-1107390 to R.G. and M.P.H. and by grants from the Brinson Foundation. 
J.M.C. is supported by NSF grant AST-1211683.
This research made use of APLpy, an open-source plotting package for Python hosted at http://aplpy.github.com; Astropy, a community-developed core Python package for Astronomy (Astropy Collaboration, 2013);  the NASA/IPAC Extragalactic Database (NED) which is operated by the Jet Propulsion Laboratory, California Institute of Technology, under contract with the National Aeronautics and Space Administration; and NASA's Astrophysics Data System.

\end{acknowledgements}


\bibliographystyle{aa}
\bibliography{refs}

\begin{thebibliography}{60}
\expandafter\ifx\csname natexlab\endcsname\relax\def\natexlab#1{#1}\fi

\bibitem[{{Adams}(2014)}]{2014PhDT........16A}
{Adams}, E.~A.~K. 2014, PhD thesis, Cornell University

\bibitem[{{Adams} {et~al.}(2015){Adams}, {Faerman}, {Janesh}, {Janowiecki},
  {Oosterloo}, {Rhode}, {Giovanelli}, {Haynes}, {Salzer}, {Sternberg},
  {Cannon}, \& {Mu{\~n}oz}}]{2015A&A...573L...3A}
{Adams}, E.~A.~K., {Faerman}, Y., {Janesh}, W.~F., {et~al.} 2015, \aap, 573, L3

\bibitem[{{Adams} {et~al.}(2013){Adams}, {Giovanelli}, \&
  {Haynes}}]{2013ApJ...768...77A}
{Adams}, E.~A.~K., {Giovanelli}, R., \& {Haynes}, M.~P. 2013, \apj, 768, 77

\bibitem[{{Beccari} {et~al.}(2016){Beccari}, {Bellazzini}, {Battaglia},
  {Ibata}, {Martin}, {Testa}, {Cignoni}, \& {Correnti}}]{2016A&A...591A..56B}
{Beccari}, G., {Bellazzini}, M., {Battaglia}, G., {et~al.} 2016, \aap, 591, A56

\bibitem[{{Bellazzini} {et~al.}(2015){Bellazzini}, {Beccari}, {Battaglia},
  {Martin}, {Testa}, {Ibata}, {Correnti}, {Cusano}, \&
  {Sani}}]{2015A&A...575A.126B}
{Bellazzini}, M., {Beccari}, G., {Battaglia}, G., {et~al.} 2015, \aap, 575,
  A126

\bibitem[{{Bernstein-Cooper} {et~al.}(2014){Bernstein-Cooper}, {Cannon},
  {Elson}, {Warren}, {Chengular}, {Skillman}, {Adams}, {Bolatto}, {Giovanelli},
  {Haynes}, {McQuinn}, {Pardy}, {Rhode}, \& {Salzer}}]{2014AJ....148...35B}
{Bernstein-Cooper}, E.~Z., {Cannon}, J.~M., {Elson}, E.~C., {et~al.} 2014, \aj,
  148, 35

\bibitem[{{Boylan-Kolchin} {et~al.}(2012){Boylan-Kolchin}, {Bullock}, \&
  {Kaplinghat}}]{2012MNRAS.422.1203B}
{Boylan-Kolchin}, M., {Bullock}, J.~S., \& {Kaplinghat}, M. 2012, \mnras, 422,
  1203

\bibitem[{{Br{\"u}ns} {et~al.}(2001){Br{\"u}ns}, {Kerp}, \&
  {Pagels}}]{2001A&A...370L..26B}
{Br{\"u}ns}, C., {Kerp}, J., \& {Pagels}, A. 2001, \aap, 370, L26

\bibitem[{{Br{\"u}ns} \& {Westmeier}(2004)}]{2004A&A...426L...9B}
{Br{\"u}ns}, C. \& {Westmeier}, T. 2004, \aap, 426, L9

\bibitem[{{Busha} {et~al.}(2010){Busha}, {Alvarez}, {Wechsler}, {Abel}, \&
  {Strigari}}]{2010ApJ...710..408B}
{Busha}, M.~T., {Alvarez}, M.~A., {Wechsler}, R.~H., {Abel}, T., \& {Strigari},
  L.~E. 2010, \apj, 710, 408

\bibitem[{{Carignan} {et~al.}(2016){Carignan}, {Libert}, {Lucero},
  {Randriamampandry}, {Jarrett}, {Oosterloo}, \&
  {Tollerud}}]{2016A&A...587L...3C}
{Carignan}, C., {Libert}, Y., {Lucero}, D.~M., {et~al.} 2016, \aap, 587, L3

\bibitem[{{Dall'Ora} {et~al.}(2006){Dall'Ora}, {Clementini}, {Kinemuchi},
  {Ripepi}, {Marconi}, {Di Fabrizio}, {Greco}, {Rodgers}, {Kuehn}, \&
  {Smith}}]{2006ApJ...653L.109D}
{Dall'Ora}, M., {Clementini}, G., {Kinemuchi}, K., {et~al.} 2006, \apjl, 653,
  L109

\bibitem[{{de Blok} {et~al.}(2008){de Blok}, {Walter}, {Brinks},
  {Trachternach}, {Oh}, \& {Kennicutt}}]{2008AJ....136.2648D}
{de Blok}, W.~J.~G., {Walter}, F., {Brinks}, E., {et~al.} 2008, \aj, 136, 2648

\bibitem[{{de Heij} {et~al.}(2002){de Heij}, {Braun}, \&
  {Burton}}]{2002A&A...391...67D}
{de Heij}, V., {Braun}, R., \& {Burton}, W.~B. 2002, \aap, 391, 67

\bibitem[{{Donovan Meyer} {et~al.}(2015){Donovan Meyer}, {Peek}, {Putman}, \&
  {Grcevich}}]{2015ApJ...808..136D}
{Donovan Meyer}, J., {Peek}, J.~E.~G., {Putman}, M., \& {Grcevich}, J. 2015,
  \apj, 808, 136

\bibitem[{{Faerman} {et~al.}(2013){Faerman}, {Sternberg}, \&
  {McKee}}]{2013ApJ...777..119F}
{Faerman}, Y., {Sternberg}, A., \& {McKee}, C.~F. 2013, \apj, 777, 119

\bibitem[{{Faridani} {et~al.}(2014){Faridani}, {Fl{\"o}er}, {Kerp}, \&
  {Westmeier}}]{2014A&A...563A..99F}
{Faridani}, S., {Fl{\"o}er}, L., {Kerp}, J., \& {Westmeier}, T. 2014, \aap,
  563, A99

\bibitem[{{Garrison-Kimmel} {et~al.}(2014){Garrison-Kimmel}, {Boylan-Kolchin},
  {Bullock}, \& {Lee}}]{2014MNRAS.438.2578G}
{Garrison-Kimmel}, S., {Boylan-Kolchin}, M., {Bullock}, J.~S., \& {Lee}, K.
  2014, \mnras, 438, 2578

\bibitem[{{Giovanelli} {et~al.}(2013){Giovanelli}, {Haynes}, {Adams}, {Cannon},
  {Rhode}, {Salzer}, {Skillman}, {Bernstein-Cooper}, \&
  {McQuinn}}]{2013AJ....146...15G}
{Giovanelli}, R., {Haynes}, M.~P., {Adams}, E.~A.~K., {et~al.} 2013, \aj, 146,
  15

\bibitem[{{Giovanelli} {et~al.}(2010){Giovanelli}, {Haynes}, {Kent}, \&
  {Adams}}]{2010ApJ...708L..22G}
{Giovanelli}, R., {Haynes}, M.~P., {Kent}, B.~R., \& {Adams}, E.~A.~K. 2010,
  \apjl, 708, L22

\bibitem[{{Giovanelli} {et~al.}(2005){Giovanelli}, {Haynes}, {Kent},
  {Perillat}, {Saintonge}, {Brosch}, {Catinella}, {Hoffman}, {Stierwalt},
  {Spekkens}, {Lerner}, {Masters}, {Momjian}, {Rosenberg}, {Springob},
  {Boselli}, {Charmandaris}, {Darling}, {Davies}, {Garcia Lambas}, {Gavazzi},
  {Giovanardi}, {Hardy}, {Hunt}, {Iovino}, {Karachentsev}, {Karachentseva},
  {Koopmann}, {Marinoni}, {Minchin}, {Muller}, {Putman}, {Pantoja}, {Salzer},
  {Scodeggio}, {Skillman}, {Solanes}, {Valotto}, {van Driel}, \& {van
  Zee}}]{2005AJ....130.2598G}
{Giovanelli}, R., {Haynes}, M.~P., {Kent}, B.~R., {et~al.} 2005, \aj, 130, 2598

\bibitem[{{Giovanelli} {et~al.}(2007){Giovanelli}, {Haynes}, {Kent},
  {Saintonge}, {Stierwalt}, {Altaf}, {Balonek}, {Brosch}, {Brown}, {Catinella},
  {Furniss}, {Goldstein}, {Hoffman}, {Koopmann}, {Kornreich}, {Mahmood},
  {Martin}, {Masters}, {Mitschang}, {Momjian}, {Nair}, {Rosenberg}, \&
  {Walsh}}]{2007AJ....133.2569G}
{Giovanelli}, R., {Haynes}, M.~P., {Kent}, B.~R., {et~al.} 2007, \aj, 133, 2569

\bibitem[{{Hoeft} {et~al.}(2006){Hoeft}, {Yepes}, {Gottl{\"o}ber}, \&
  {Springel}}]{2006MNRAS.371..401H}
{Hoeft}, M., {Yepes}, G., {Gottl{\"o}ber}, S., \& {Springel}, V. 2006, \mnras,
  371, 401

\bibitem[{{Hoffman} {et~al.}(1996){Hoffman}, {Salpeter}, {Farhat}, {Roos},
  {Williams}, \& {Helou}}]{1996ApJS..105..269H}
{Hoffman}, G.~L., {Salpeter}, E.~E., {Farhat}, B., {et~al.} 1996, \apjs, 105,
  269

\bibitem[{{Janesh} {et~al.}(2015){Janesh}, {Rhode}, {Salzer}, {Janowiecki},
  {Adams}, {Haynes}, {Giovanelli}, {Cannon}, \&
  {Mu{\~n}oz}}]{2015ApJ...811...35J}
{Janesh}, W., {Rhode}, K.~L., {Salzer}, J.~J., {et~al.} 2015, \apj, 811, 35

\bibitem[{{Jones} {et~al.}(2016){Jones}, {Papastergis}, {Haynes}, \&
  {Giovanelli}}]{2016MNRAS.457.4393J}
{Jones}, M.~G., {Papastergis}, E., {Haynes}, M.~P., \& {Giovanelli}, R. 2016,
  \mnras, 457, 4393

\bibitem[{{Kauffmann} {et~al.}(1993){Kauffmann}, {White}, \&
  {Guiderdoni}}]{1993MNRAS.264..201K}
{Kauffmann}, G., {White}, S.~D.~M., \& {Guiderdoni}, B. 1993, \mnras, 264, 201

\bibitem[{{Klypin} {et~al.}(1999){Klypin}, {Kravtsov}, {Valenzuela}, \&
  {Prada}}]{1999ApJ...522...82K}
{Klypin}, A., {Kravtsov}, A.~V., {Valenzuela}, O., \& {Prada}, F. 1999, \apj,
  522, 82

\bibitem[{{Koch} {et~al.}(2009){Koch}, {Wilkinson}, {Kleyna}, {Irwin},
  {Zucker}, {Belokurov}, {Gilmore}, {Fellhauer}, \&
  {Evans}}]{2009ApJ...690..453K}
{Koch}, A., {Wilkinson}, M.~I., {Kleyna}, J.~T., {et~al.} 2009, \apj, 690, 453

\bibitem[{{Kravtsov}(2010)}]{2010AdAst2010E...8K}
{Kravtsov}, A. 2010, Advances in Astronomy, 2010, 281913

\bibitem[{{Martin} {et~al.}(2010){Martin}, {Papastergis}, {Giovanelli},
  {Haynes}, {Springob}, \& {Stierwalt}}]{2010ApJ...723.1359M}
{Martin}, A.~M., {Papastergis}, E., {Giovanelli}, R., {et~al.} 2010, \apj, 723,
  1359

\bibitem[{{McGaugh}(2012)}]{2012AJ....143...40M}
{McGaugh}, S.~S. 2012, \aj, 143, 40

\bibitem[{{McMullin} {et~al.}(2007){McMullin}, {Waters}, {Schiebel}, {Young},
  \& {Golap}}]{2007ASPC..376..127M}
{McMullin}, J.~P., {Waters}, B., {Schiebel}, D., {Young}, W., \& {Golap}, K.
  2007, in Astronomical Society of the Pacific Conference Series, Vol. 376,
  Astronomical Data Analysis Software and Systems XVI, ed. R.~A. {Shaw},
  F.~{Hill}, \& D.~J. {Bell}, 127

\bibitem[{{McNichols} {et~al.}(2016){McNichols}, {Teich}, {Nims}, {Cannon},
  {Adams}, {Bernstein-Cooper}, {Giovanelli}, {Haynes}, {J\'ozsa}, {McQuinn},
  {Salzer}, {Skillman}, {Warren}, {Dolphin}, {Elson}, {Haurberg}, {Ott},
  {Saintonge}, {Cave}, {Hagen}, {Huang}, {Janowiecki}, {Marshall}, {Moody},
  {Thomann}, \& {Van Sistine}}]{McNichols}
{McNichols}, A.~T., {Teich}, Y.~G., {Nims}, E., {et~al.} 2016, \apj, in press

\bibitem[{{McQuinn} {et~al.}(2015){McQuinn}, {Skillman}, {Dolphin}, {Cannon},
  {Salzer}, {Rhode}, {Adams}, {Berg}, {Giovanelli}, {Girardi}, \&
  {Haynes}}]{2015ApJ...812..158M}
{McQuinn}, K.~B.~W., {Skillman}, E.~D., {Dolphin}, A., {et~al.} 2015, \apj,
  812, 158

\bibitem[{{Moore} {et~al.}(1999){Moore}, {Ghigna}, {Governato}, {Lake},
  {Quinn}, {Stadel}, \& {Tozzi}}]{1999ApJ...524L..19M}
{Moore}, B., {Ghigna}, S., {Governato}, F., {et~al.} 1999, \apjl, 524, L19

\bibitem[{{O{\~n}orbe} {et~al.}(2015){O{\~n}orbe}, {Boylan-Kolchin}, {Bullock},
  {Hopkins}, {Kere{\v s}}, {Faucher-Gigu{\`e}re}, {Quataert}, \&
  {Murray}}]{2015MNRAS.454.2092O}
{O{\~n}orbe}, J., {Boylan-Kolchin}, M., {Bullock}, J.~S., {et~al.} 2015,
  \mnras, 454, 2092

\bibitem[{{Oh} {et~al.}(2011){Oh}, {Brook}, {Governato}, {Brinks}, {Mayer}, {de
  Blok}, {Brooks}, \& {Walter}}]{2011AJ....142...24O}
{Oh}, S.-H., {Brook}, C., {Governato}, F., {et~al.} 2011, \aj, 142, 24

\bibitem[{{Papastergis} {et~al.}(2015){Papastergis}, {Giovanelli}, {Haynes}, \&
  {Shankar}}]{2015A&A...574A.113P}
{Papastergis}, E., {Giovanelli}, R., {Haynes}, M.~P., \& {Shankar}, F. 2015,
  \aap, 574, A113

\bibitem[{{Papastergis} {et~al.}(2011){Papastergis}, {Martin}, {Giovanelli}, \&
  {Haynes}}]{2011ApJ...739...38P}
{Papastergis}, E., {Martin}, A.~M., {Giovanelli}, R., \& {Haynes}, M.~P. 2011,
  \apj, 739, 38

\bibitem[{{Roychowdhury} {et~al.}(2010){Roychowdhury}, {Chengalur}, {Begum}, \&
  {Karachentsev}}]{2010MNRAS.404L..60R}
{Roychowdhury}, S., {Chengalur}, J.~N., {Begum}, A., \& {Karachentsev}, I.~D.
  2010, \mnras, 404, L60

\bibitem[{{Ryan-Weber} {et~al.}(2008){Ryan-Weber}, {Begum}, {Oosterloo}, {Pal},
  {Irwin}, {Belokurov}, {Evans}, \& {Zucker}}]{2008MNRAS.384..535R}
{Ryan-Weber}, E.~V., {Begum}, A., {Oosterloo}, T., {et~al.} 2008, \mnras, 384,
  535

\bibitem[{{Sand} {et~al.}(2015){Sand}, {Crnojevi{\'c}}, {Bennet}, {Willman},
  {Hargis}, {Strader}, {Olszewski}, {Tollerud}, {Simon}, {Caldwell},
  {Guhathakurta}, {James}, {Koposov}, {McLeod}, {Morrell}, {Peacock},
  {Salinas}, {Seth}, {Stark}, \& {Toloba}}]{2015ApJ...806...95S}
{Sand}, D.~J., {Crnojevi{\'c}}, D., {Bennet}, P., {et~al.} 2015, \apj, 806, 95

\bibitem[{{Saul} {et~al.}(2012){Saul}, {Peek}, {Grcevich}, {Putman}, {Douglas},
  {Korpela}, {Stanimirovi{\'c}}, {Heiles}, {Gibson}, {Lee}, {Begum}, {Brown},
  {Burkhart}, {Hamden}, {Pingel}, \& {Tonnesen}}]{2012ApJ...758...44S}
{Saul}, D.~R., {Peek}, J.~E.~G., {Grcevich}, J., {et~al.} 2012, \apj, 758, 44

\bibitem[{{Sault} {et~al.}(1995){Sault}, {Teuben}, \&
  {Wright}}]{1995ASPC...77..433S}
{Sault}, R.~J., {Teuben}, P.~J., \& {Wright}, M.~C.~H. 1995, in Astronomical
  Society of the Pacific Conference Series, Vol.~77, Astronomical Data Analysis
  Software and Systems IV, ed. R.~A. {Shaw}, H.~E. {Payne}, \& J.~J.~E.
  {Hayes}, 433

\bibitem[{{Sawala} {et~al.}(2016){Sawala}, {Frenk}, {Fattahi}, {Navarro},
  {Bower}, {Crain}, {Dalla Vecchia}, {Furlong}, {Helly}, {Jenkins}, {Oman},
  {Schaller}, {Schaye}, {Theuns}, {Trayford}, \& {White}}]{2016MNRAS.457.1931S}
{Sawala}, T., {Frenk}, C.~S., {Fattahi}, A., {et~al.} 2016, \mnras, 457, 1931

\bibitem[{{Sawala} {et~al.}(2015){Sawala}, {Frenk}, {Fattahi}, {Navarro},
  {Bower}, {Crain}, {Dalla Vecchia}, {Furlong}, {Jenkins}, {McCarthy}, {Qu},
  {Schaller}, {Schaye}, \& {Theuns}}]{2015MNRAS.448.2941S}
{Sawala}, T., {Frenk}, C.~S., {Fattahi}, A., {et~al.} 2015, \mnras, 448, 2941

\bibitem[{{Simpson} {et~al.}(2013){Simpson}, {Bryan}, {Johnston}, {Smith}, {Mac
  Low}, {Sharma}, \& {Tumlinson}}]{2013MNRAS.432.1989S}
{Simpson}, C.~M., {Bryan}, G.~L., {Johnston}, K.~V., {et~al.} 2013, \mnras,
  432, 1989

\bibitem[{{Spekkens} {et~al.}(2014){Spekkens}, {Urbancic}, {Mason}, {Willman},
  \& {Aguirre}}]{2014ApJ...795L...5S}
{Spekkens}, K., {Urbancic}, N., {Mason}, B.~S., {Willman}, B., \& {Aguirre},
  J.~E. 2014, \apjl, 795, L5

\bibitem[{{Teich} {et~al.}(2016){Teich}, {McNichols}, {Nims}, {Cannon},
  {Adams}, {Giovanelli}, {Haynes}, {McQuinn}, {Salzer}, {Skillman},
  {Bernstein-Cooper}, {Dolphin}, {Elson}, {Haurberg}, {J\'ozsa}, {Ott},
  {Saintonge}, {Warren}, {Cave}, {Hagen}, {Huang}, {Janowiecki}, {Marshall},
  {Moody}, {Thomann}, \& {Van Sistine}}]{Teich}
{Teich}, Y.~G., {McNichols}, A.~T., {Nims}, E., {et~al.} 2016, \apj, in press

\bibitem[{{Tully} {et~al.}(2013){Tully}, {Courtois}, {Dolphin}, {Fisher},
  {H{\'e}raudeau}, {Jacobs}, {Karachentsev}, {Makarov}, {Makarova},
  {Mitronova}, {Rizzi}, {Shaya}, {Sorce}, \& {Wu}}]{2013AJ....146...86T}
{Tully}, R.~B., {Courtois}, H.~M., {Dolphin}, A.~E., {et~al.} 2013, \aj, 146,
  86

\bibitem[{{Vandenbroucke} {et~al.}(2016){Vandenbroucke}, {Verbeke}, \& {De
  Rijcke}}]{2016MNRAS.458..912V}
{Vandenbroucke}, B., {Verbeke}, R., \& {De Rijcke}, S. 2016, \mnras, 458, 912

\bibitem[{{Vogelsberger} {et~al.}(2014){Vogelsberger}, {Genel}, {Springel},
  {Torrey}, {Sijacki}, {Xu}, {Snyder}, {Bird}, {Nelson}, \&
  {Hernquist}}]{2014Natur.509..177V}
{Vogelsberger}, M., {Genel}, S., {Springel}, V., {et~al.} 2014, \nat, 509, 177

\bibitem[{{Wakker} \& {van Woerden}(1991)}]{1991A&A...250..509W}
{Wakker}, B.~P. \& {van Woerden}, H. 1991, \aap, 250, 509

\bibitem[{{Walker} \& {Pe{\~n}arrubia}(2011)}]{2011ApJ...742...20W}
{Walker}, M.~G. \& {Pe{\~n}arrubia}, J. 2011, \apj, 742, 20

\bibitem[{{Weisz} {et~al.}(2012){Weisz}, {Zucker}, {Dolphin}, {Martin}, {de
  Jong}, {Holtzman}, {Dalcanton}, {Gilbert}, {Williams}, {Bell}, {Belokurov},
  \& {Wyn Evans}}]{2012ApJ...748...88W}
{Weisz}, D.~R., {Zucker}, D.~B., {Dolphin}, A.~E., {et~al.} 2012, \apj, 748, 88

\bibitem[{{Westmeier} {et~al.}(2005{\natexlab{a}}){Westmeier}, {Braun}, \&
  {Thilker}}]{2005A&A...436..101W}
{Westmeier}, T., {Braun}, R., \& {Thilker}, D. 2005{\natexlab{a}}, \aap, 436,
  101

\bibitem[{{Westmeier} {et~al.}(2005{\natexlab{b}}){Westmeier}, {Br{\"u}ns}, \&
  {Kerp}}]{2005A&A...432..937W}
{Westmeier}, T., {Br{\"u}ns}, C., \& {Kerp}, J. 2005{\natexlab{b}}, \aap, 432,
  937

\bibitem[{{Wetzel} {et~al.}(2016){Wetzel}, {Hopkins}, {Kim}, {Faucher-Giguere},
  {Keres}, \& {Quataert}}]{2016arXiv160205957W}
{Wetzel}, A.~R., {Hopkins}, P.~F., {Kim}, J.-h., {et~al.} 2016, ArXiv e-prints

\bibitem[{{Wheeler} {et~al.}(2015){Wheeler}, {O{\~n}orbe}, {Bullock},
  {Boylan-Kolchin}, {Elbert}, {Garrison-Kimmel}, {Hopkins}, \& {Kere{\v
  s}}}]{2015MNRAS.453.1305W}
{Wheeler}, C., {O{\~n}orbe}, J., {Bullock}, J.~S., {et~al.} 2015, \mnras, 453,
  1305

\end{thebibliography}


\appendix
\section{Presentation of the \hi\ data}\label{sec:sources}

\begin{figure*}
\centering
\includegraphics[keepaspectratio,width=0.6\linewidth]{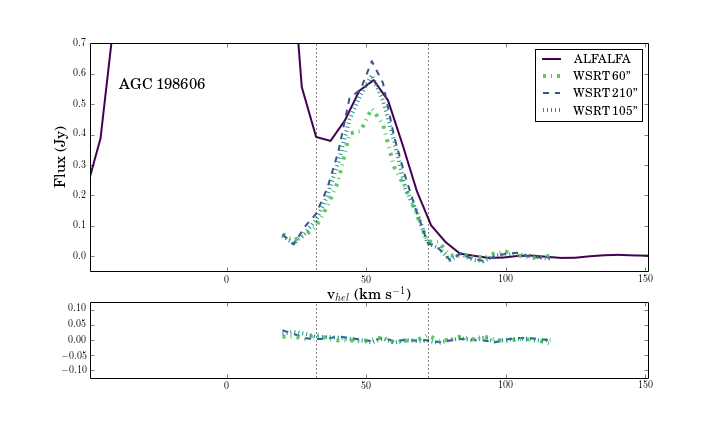}
\includegraphics[keepaspectratio,width=0.35\linewidth]{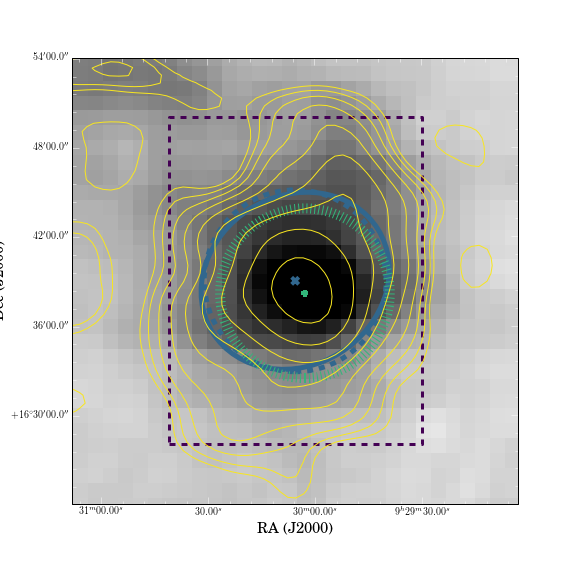}
\includegraphics[keepaspectratio,height=.5\linewidth]{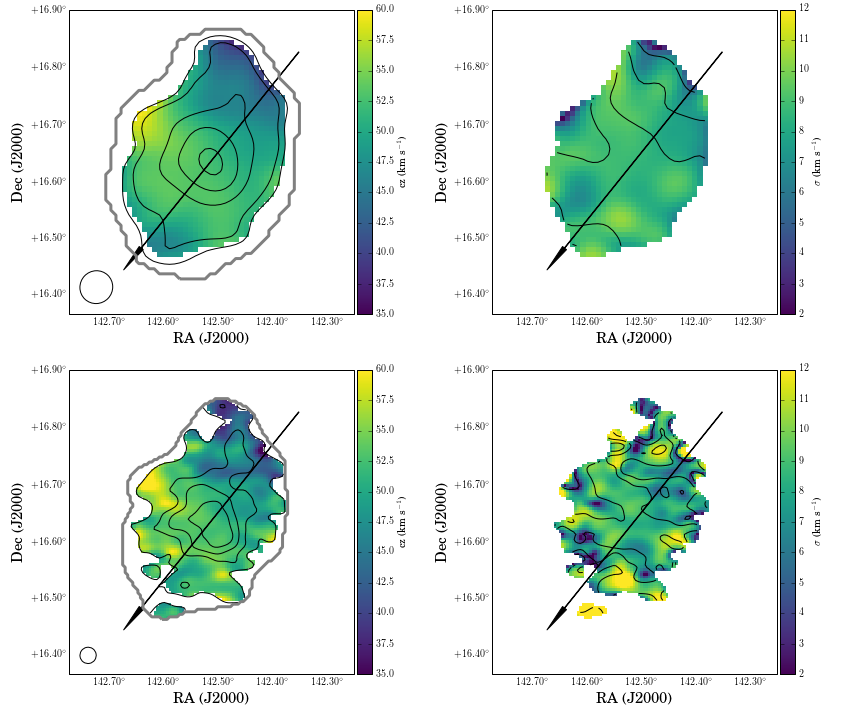}
\includegraphics[keepaspectratio,height=.5\linewidth,clip=true,trim=0cm 2cm 0cm 2.5cm]{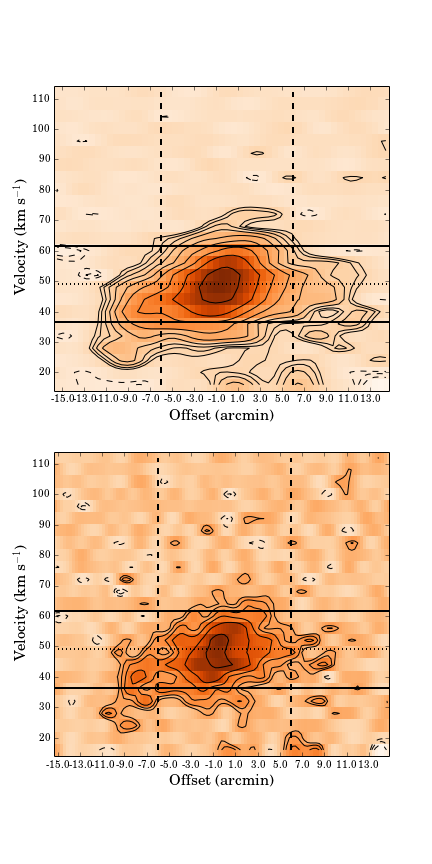}
\includegraphics[keepaspectratio,height=.25\linewidth]{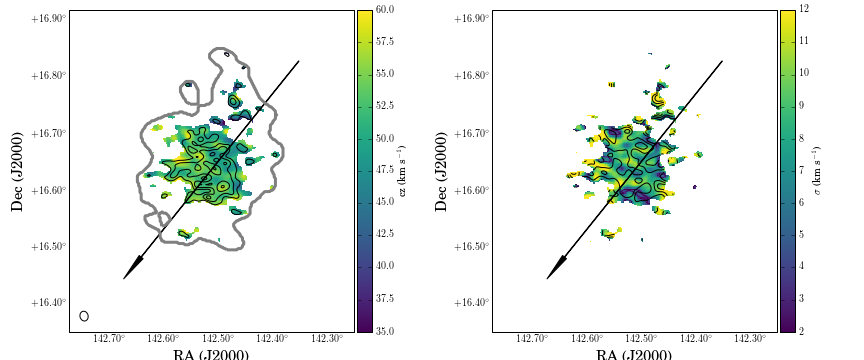}
\includegraphics[keepaspectratio,height=.25\linewidth,clip=true,trim=0cm 1cm 0cm 1.25cm]{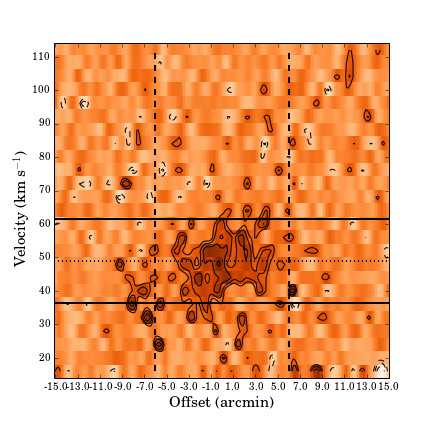}
\caption{ALFALFA and WSRT data for AGC\,198606.
{\it Top, left}: Spectra for the source: solid purple is the ALFALFA spectrum, dashed blue-purple the 210\arcsec\ WSRT spectrum,
 dotted blue the 105\arcsec\ WSRT spectrum, and the dashed-dotted green the 60\arcsec\ WSRT spectrum. 
 The light dotted black lines indicate the velocity range used
for creation of the WSRT moment maps. 
The bottom panel shows the noise from a representative region of the WSRT data cube.
{\it Top, right}: The ALFALFA total intensity \hi\ map in grayscale  with the region used for source measurement 
highlighted by the dashed  rectangle. The yellow contours are the WSRT 210\arcsec\ moment zero map 
contours starting from 2 $\times$ rms and increasing by $\sqrt{2}$; dashed negative contours with the same
spacing are also shown. 
The blue ellipse represents the ALFALFA fitted ellipse and the dashed blue circle the equivalent $\bar a$, 
average angular size. The dotted green circle represents the effective half-flux aperture derived in this work.
{\it Bottom}: The WSRT \hi\ data. From left to right, the panels show the velocity field with column density contours overlaid 
(and the mask used to extract the integrated flux density and spectrum in dark gray),
the velocity dispersion map with lines of constant velocity from the velocity field, and a position velocity slice along
the arrow shown in the left and central panels. The \hi\ column density contours for the three resolutions are  [0.7,1, 2, 3, 4] $\times 10^{19}$, [1, 2, 3, 4, 5] $\times 10^{19}$ 
 and [3, 4, 5, 6] $\times 10^{19}$ atoms cm$^{-2}$.  
 The constant velocity contours are [42, 46, 50, 54, 60] \kms.
 Contours for the position-velocity slice start at 2 $\times$ rms of the data cube
 and increase by $\sqrt{2}$; negative contours with the same spacing are also shown.
}
\label{fig:hvc214.78+42.45+47}
\end{figure*}

\begin{figure*}
\centering
\includegraphics[keepaspectratio,width=0.6\linewidth]{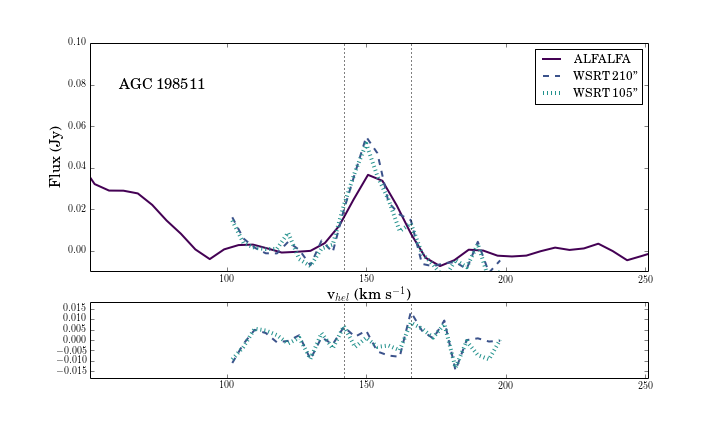}
\includegraphics[keepaspectratio,width=0.35\linewidth]{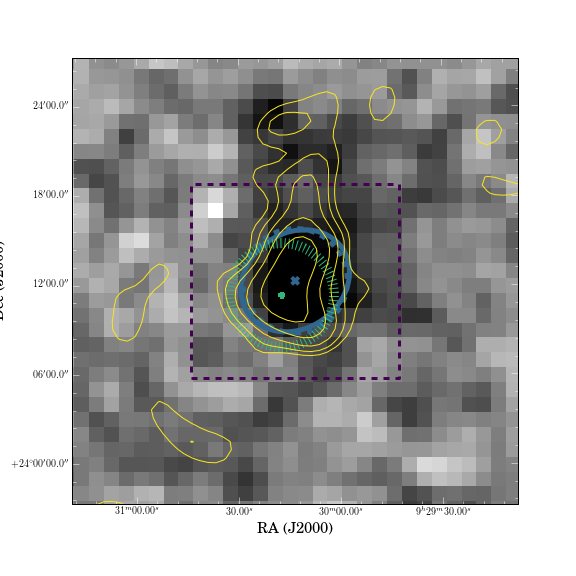}
\includegraphics[keepaspectratio,height=.5\linewidth]{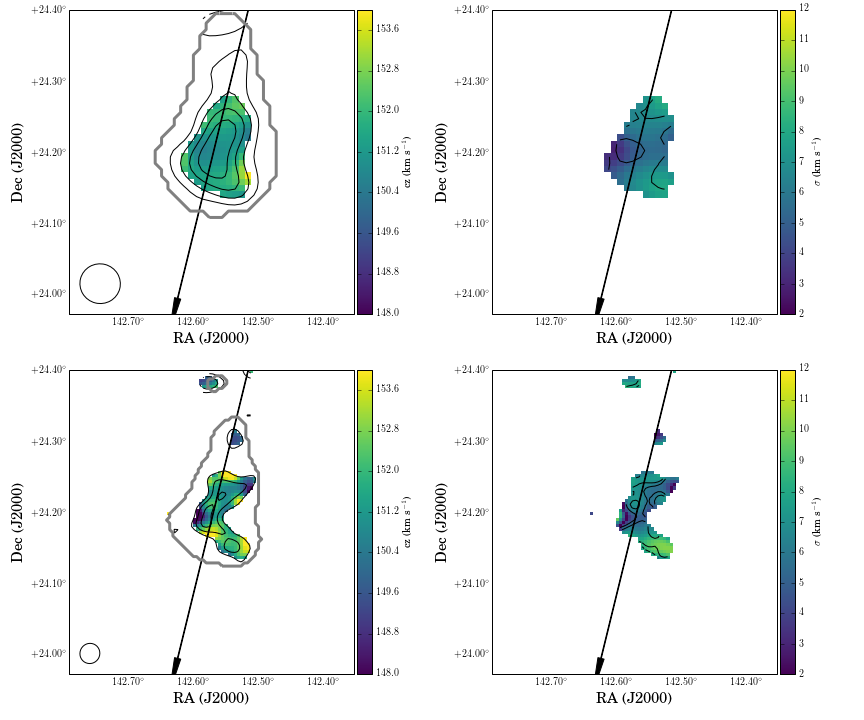}
\includegraphics[keepaspectratio,height=.5\linewidth,clip=true,trim=0cm 2cm 0cm 2.5cm]{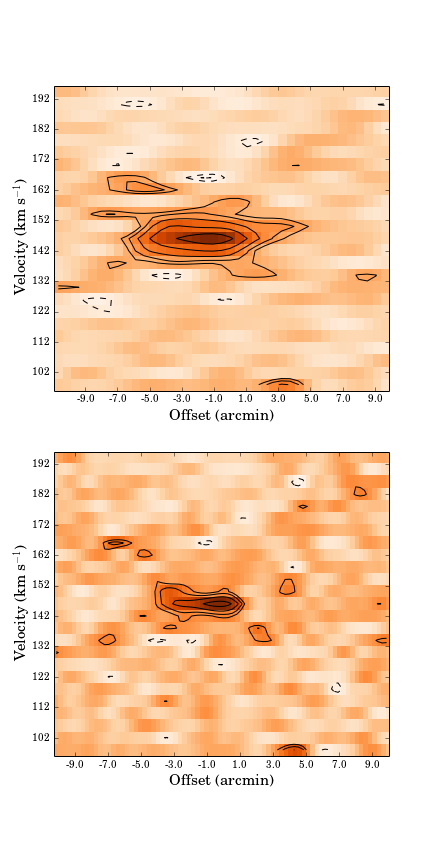}
\caption{ALFALFA and WSRT data for AGC\,198511 as in Figure \ref{fig:hvc214.78+42.45+47} but with \hi\ column density contours at  [2, 3, 4, 5] $\times 10^{18}$ and [5, 7, 9, 12.5] $\times 10^{18}$ atoms cm$^{-2}$ for the 210\arcsec\ and 105\arcsec\ data and velocity contours at [151, 152] \kms.}
\label{fig:hvc204.88+44.86+147}
\end{figure*}

\begin{figure*}
\centering
\includegraphics[keepaspectratio,width=0.6\linewidth]{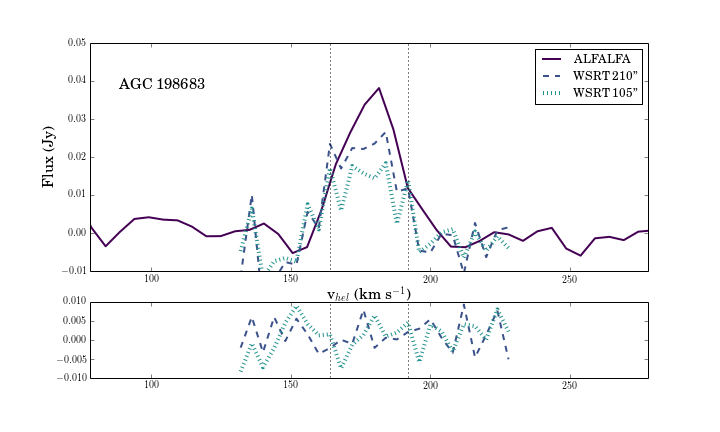}
\includegraphics[keepaspectratio,width=0.35\linewidth]{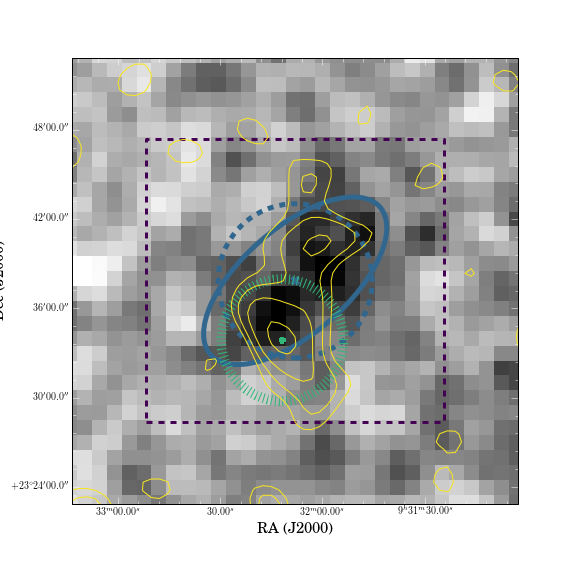}
\includegraphics[keepaspectratio,height=.5\linewidth]{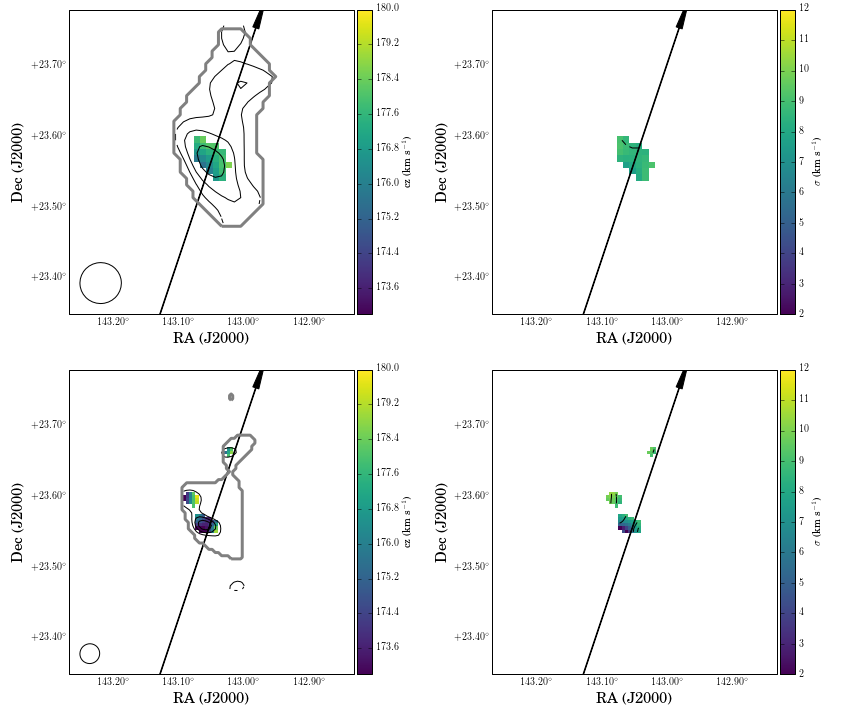}
\includegraphics[keepaspectratio,height=.5\linewidth,clip=true,trim=0cm 2cm 0cm 2.5cm]{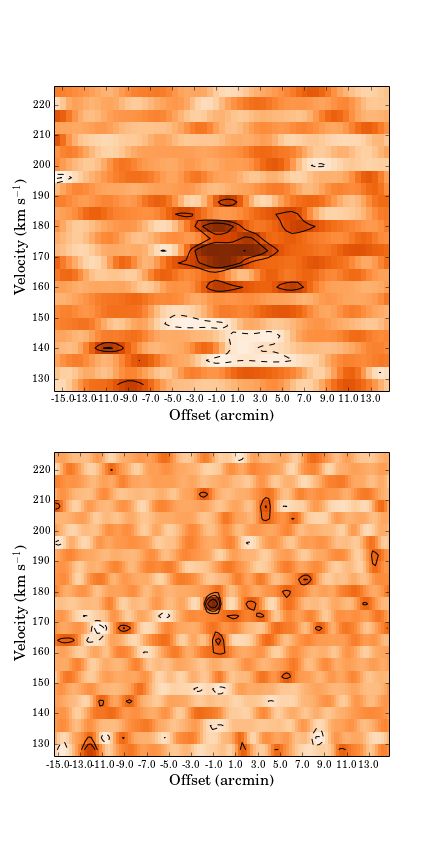}
\caption{ALFALFA and WSRT data for AGC\,198683 as in Figure \ref{fig:hvc214.78+42.45+47} but with \hi\ column density contours at  [2, 3, 4] $\times 10^{18}$ and [6, 8, 10] $\times 10^{18}$ atoms cm$^{-2}$ for the 210\arcsec\ and 105\arcsec\ data and velocity contours at [175, 178, 181] \kms.}
\label{fig:hvc205.83+45.14+173}
\end{figure*}

\begin{figure*}
\centering
\includegraphics[keepaspectratio,width=0.6\linewidth]{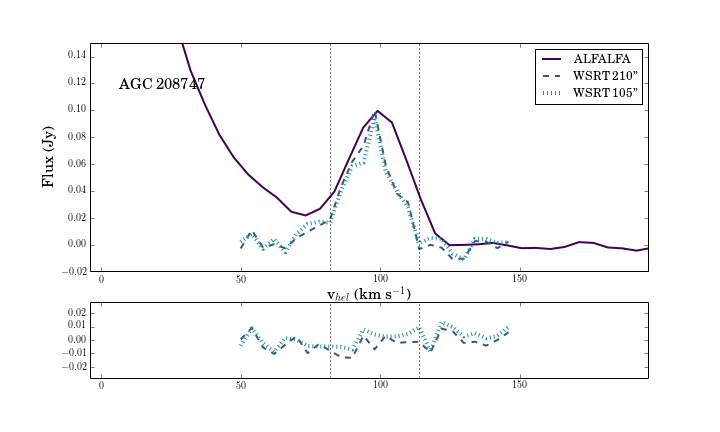}
\includegraphics[keepaspectratio,width=0.35\linewidth]{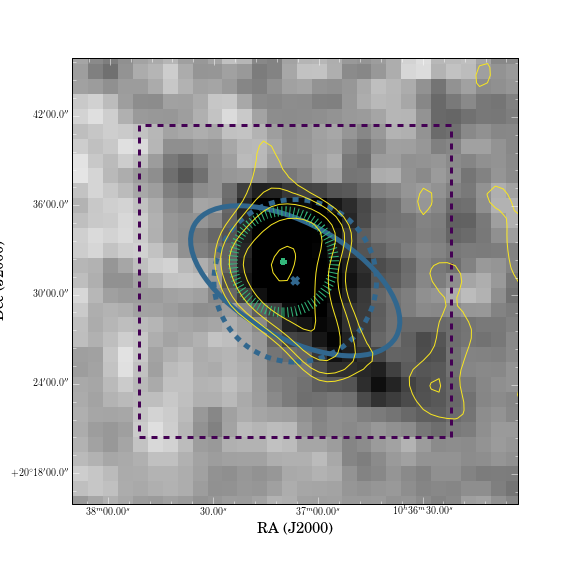}
\includegraphics[keepaspectratio,height=.5\linewidth]{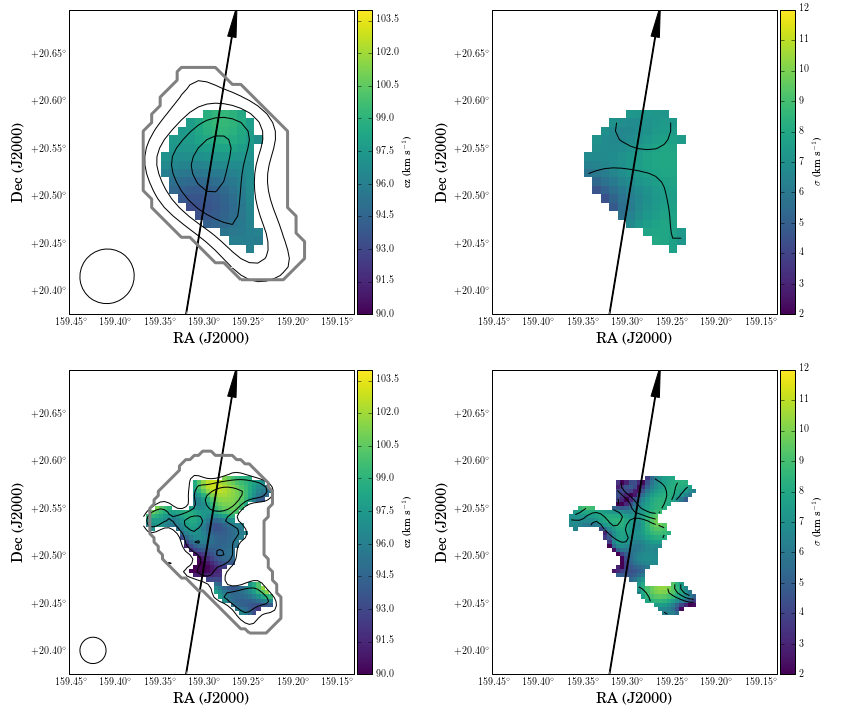}
\includegraphics[keepaspectratio,height=.5\linewidth,clip=true,trim=0cm 2cm 0cm 2.5cm]{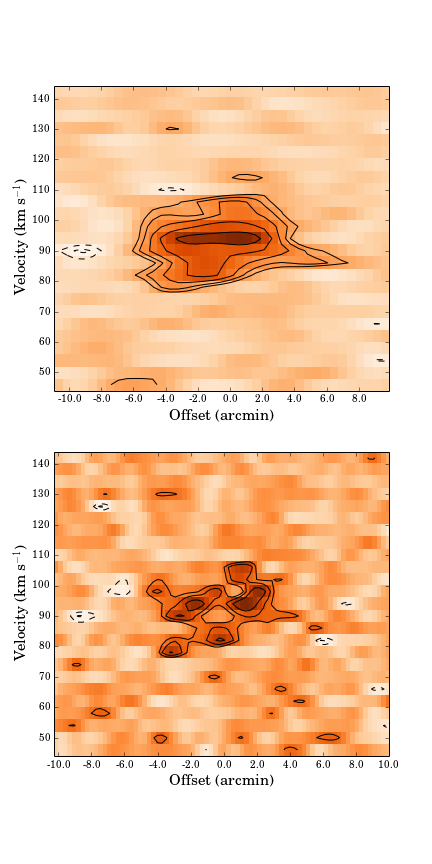}
\caption{ALFALFA and WSRT data for AGC\,208747 as in Figure \ref{fig:hvc214.78+42.45+47} but with \hi\ column density contours at  [4, 6, 8,10] $\times 10^{18}$ and [8, 10, 12.5] $\times 10^{18}$ atoms cm$^{-2}$ for the 210\arcsec\ and 105\arcsec\ data and velocity contours at [96, 98, 100] \kms.}
\label{fig:hvc217.77+58.67+96}
\end{figure*}

\begin{figure*}
\centering
\includegraphics[keepaspectratio,width=0.6\linewidth]{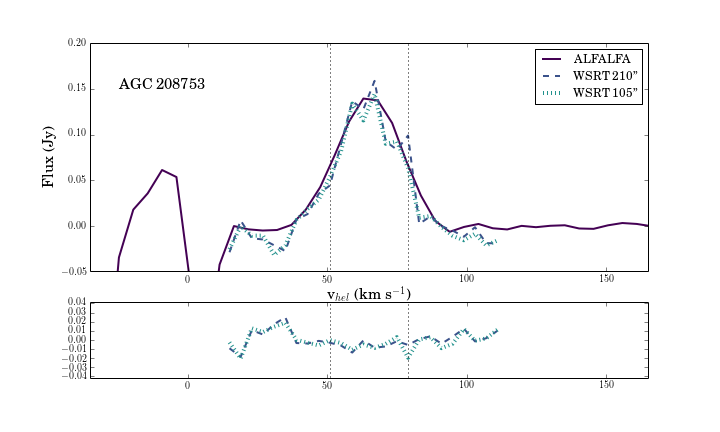}
\includegraphics[keepaspectratio,width=0.35\linewidth]{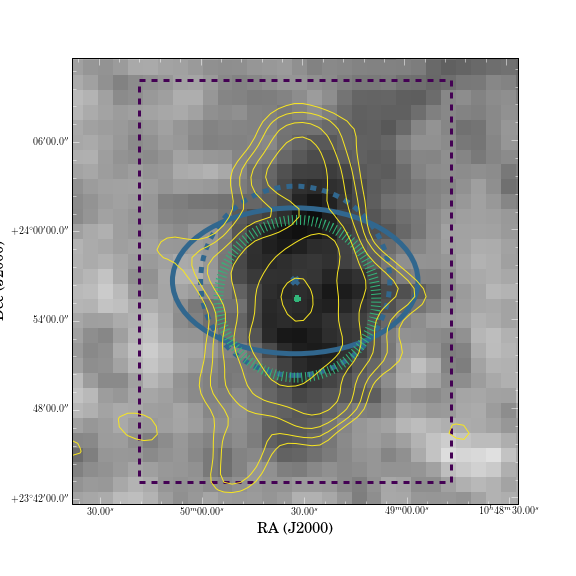}
\includegraphics[keepaspectratio,height=.5\linewidth]{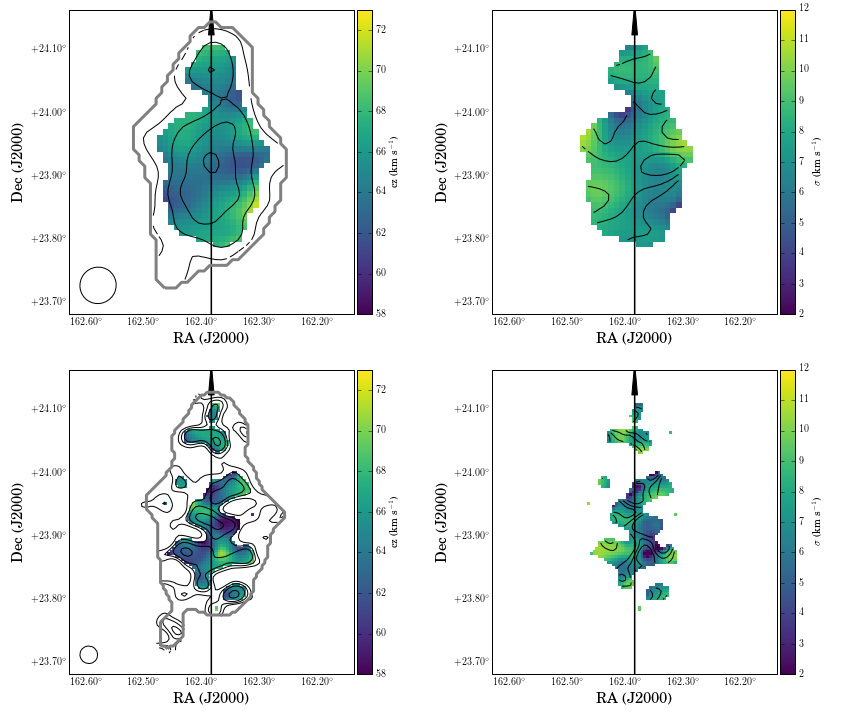}
\includegraphics[keepaspectratio,height=.5\linewidth,clip=true,trim=0cm 2cm 0cm 2.5cm]{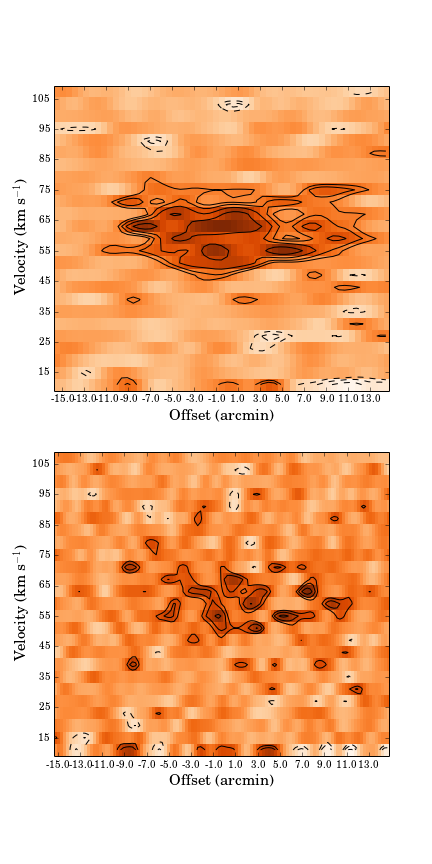}
\caption{ALFALFA and WSRT data for AGC\,208753 as in Figure \ref{fig:hvc214.78+42.45+47} but with \hi\ column density contours at  [3, 5, 7, 9] $\times 10^{18}$ and [5, 7, 9, 12.5] $\times 10^{18}$ atoms cm$^{-2}$ for the 210\arcsec\ and 105\arcsec\ data and velocity contours at [63, 65, 67] \kms.}
\label{fig:hvc212.68+62.39+64}
\end{figure*}

\begin{figure*}
\centering
\includegraphics[keepaspectratio,width=0.6\linewidth]{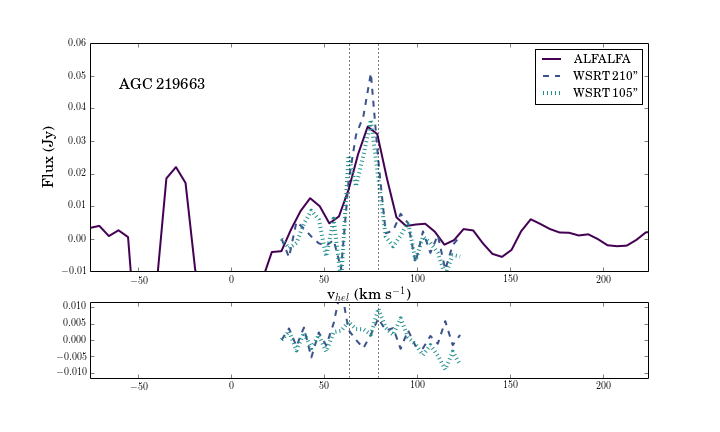}
\includegraphics[keepaspectratio,width=0.35\linewidth]{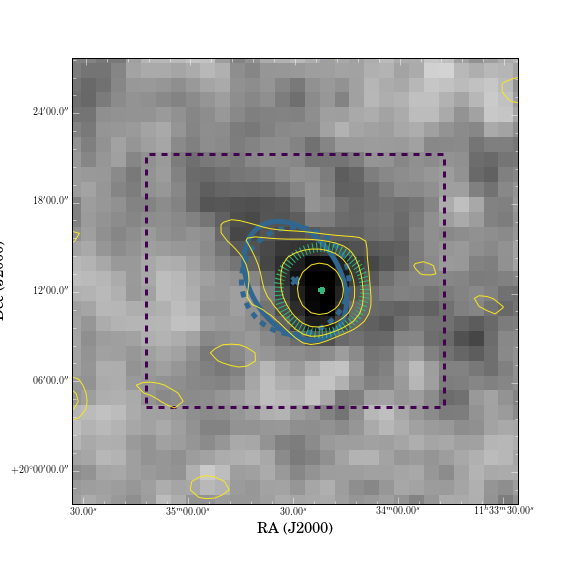}
\includegraphics[keepaspectratio,height=.5\linewidth]{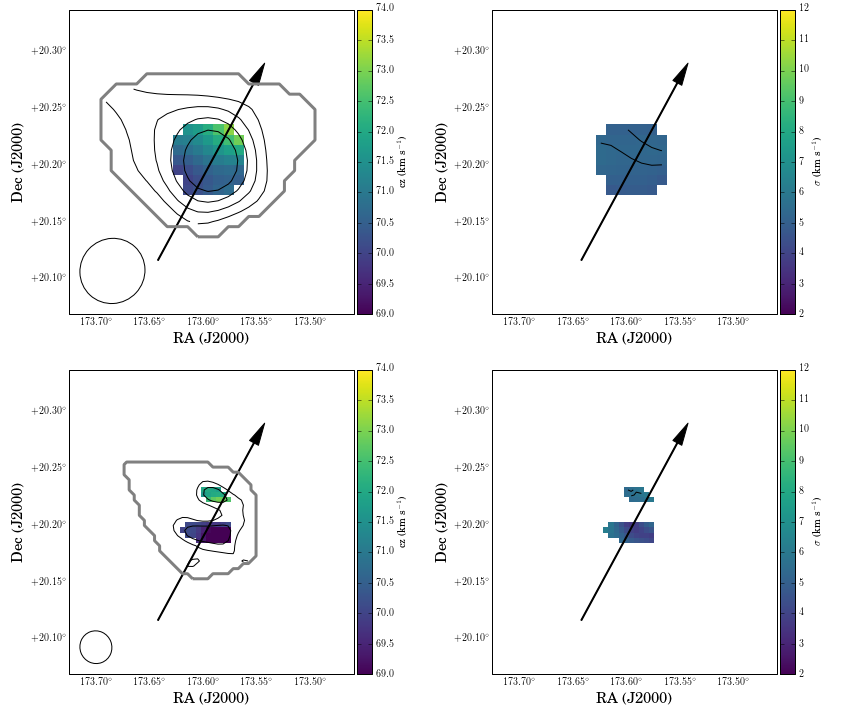}
\includegraphics[keepaspectratio,height=.5\linewidth,clip=true,trim=0cm 2cm 0cm 2.5cm]{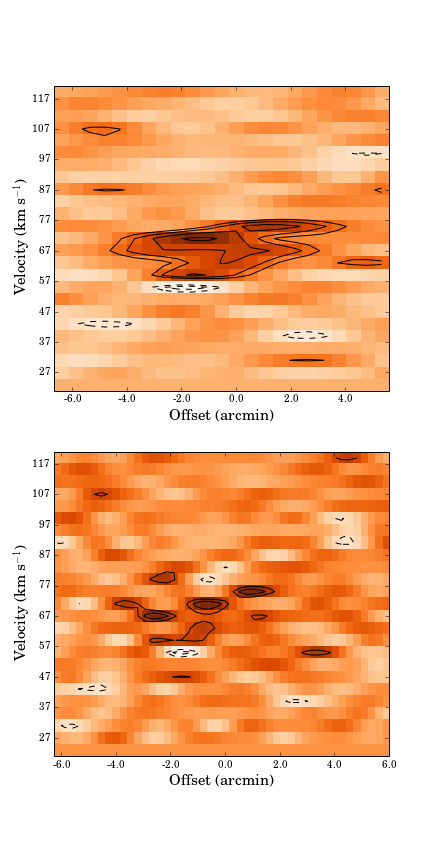}
\caption{ALFALFA and WSRT data for AGC\,219663 as in Figure \ref{fig:hvc214.78+42.45+47} but with \hi\ column density contours at  [2.5, 3.5, 4.5, 5.5] $\times 10^{18}$ and [7, 9] $\times 10^{18}$ atoms cm$^{-2}$ for the 210\arcsec\ and 105\arcsec\ data and velocity contours at [71,72] \kms.}
\label{fig:hvc230.27+71.10+76}
\end{figure*}

\begin{figure*}
\centering
\includegraphics[keepaspectratio,width=0.6\linewidth]{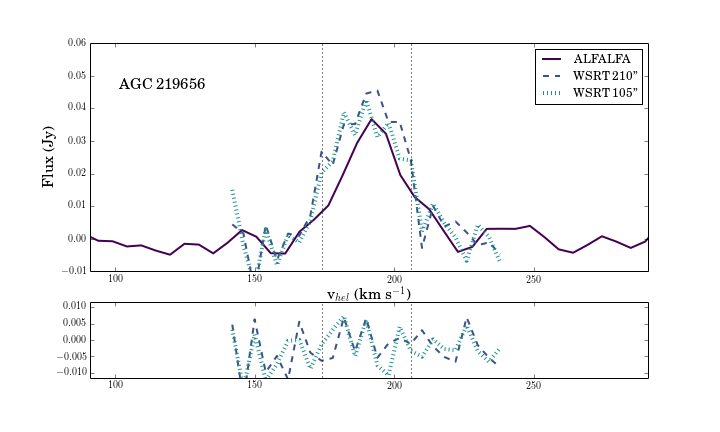}
\includegraphics[keepaspectratio,width=0.35\linewidth]{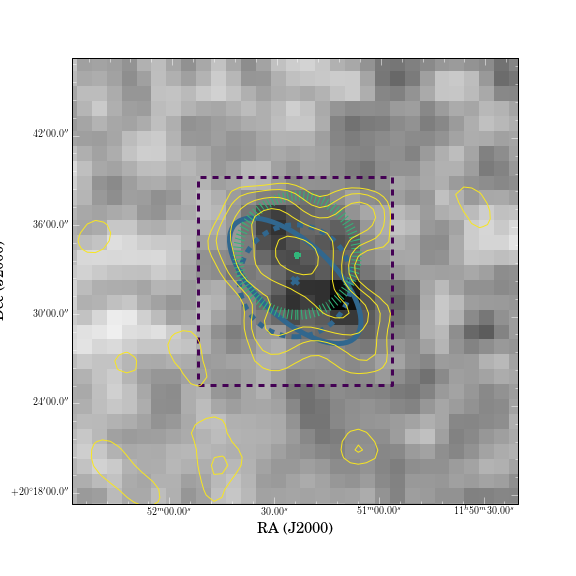}
\includegraphics[keepaspectratio,height=.5\linewidth]{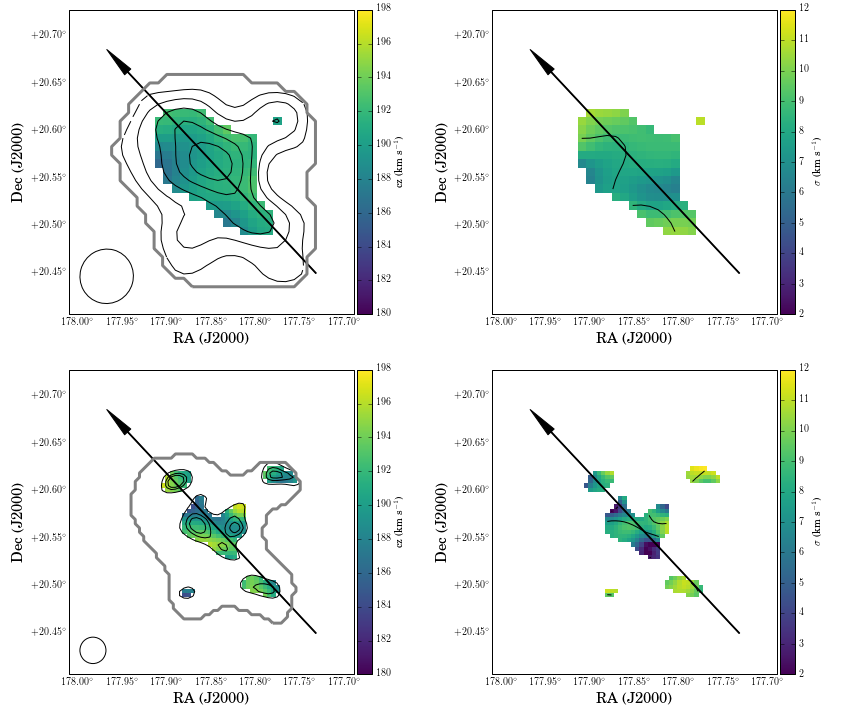}
\includegraphics[keepaspectratio,height=.5\linewidth,clip=true,trim=0cm 2cm 0cm 2.5cm]{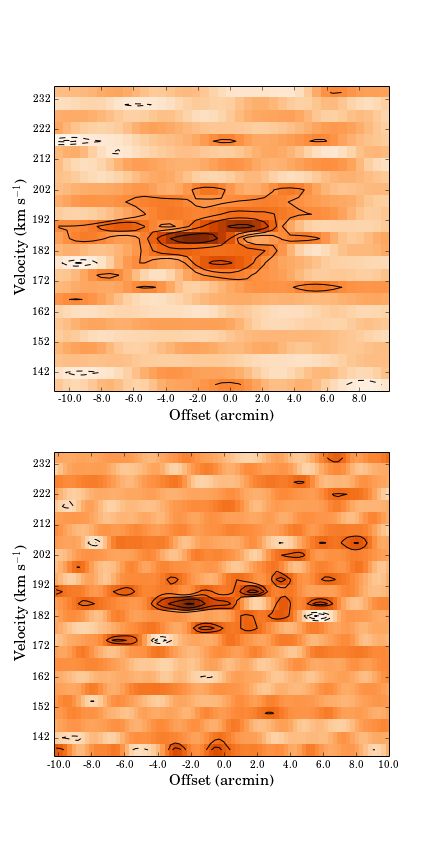}
\caption{ALFALFA and WSRT data for AGC\,219656 as in Figure \ref{fig:hvc214.78+42.45+47} but with \hi\ column density contours at  [2, 3, 4, 5, 6] $\times 10^{18}$ and [7, 9, 10] $\times 10^{18}$ atoms cm$^{-2}$ for the 210\arcsec\ and 105\arcsec\ data and velocity contours at [185, 190] \kms.}
\label{fig:hvc235.38+74.79+195}
\end{figure*}

\begin{figure*}
\centering
\includegraphics[keepaspectratio,width=0.6\linewidth]{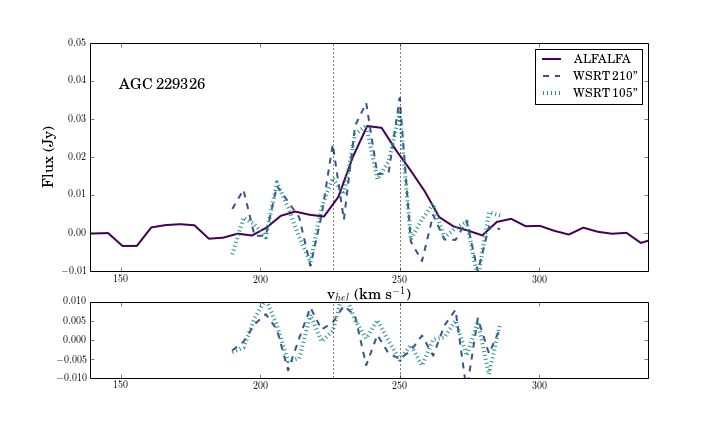}
\includegraphics[keepaspectratio,width=0.35\linewidth]{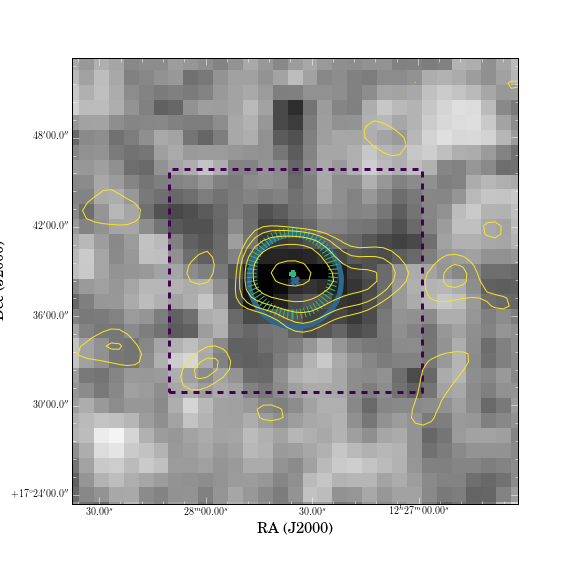}
\includegraphics[keepaspectratio,height=.5\linewidth]{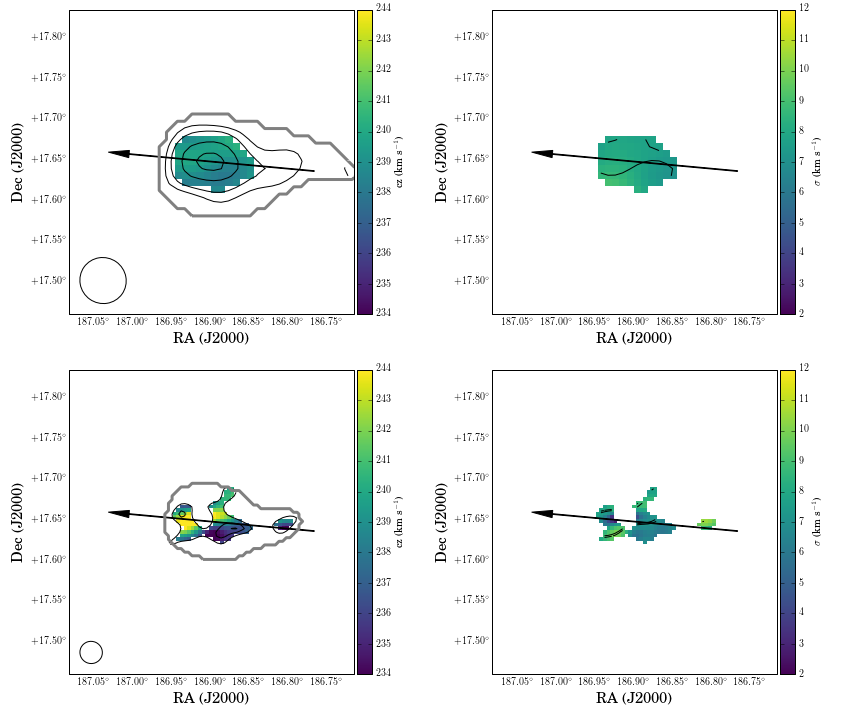}
\includegraphics[keepaspectratio,height=.5\linewidth,clip=true,trim=0cm 2cm 0cm 2.5cm]{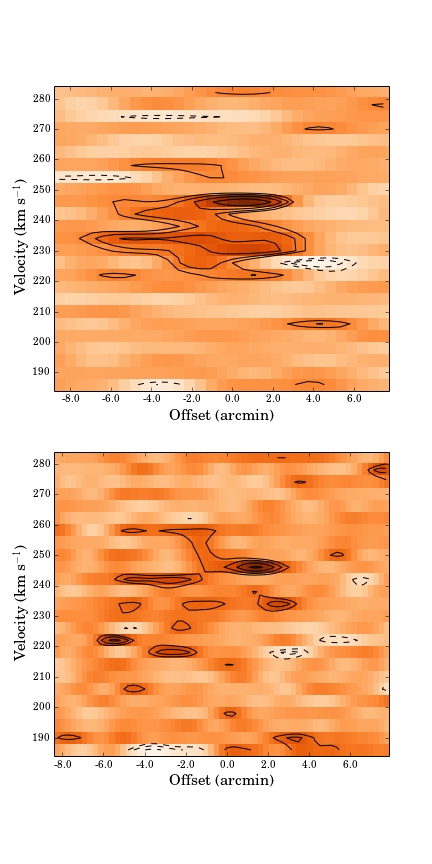}
\caption{
ALFALFA and WSRT data for AGC\,229326 as in Figure  \ref{fig:hvc214.78+42.45+47}
but with \hi\ column density contours at [2.5, 3.5, 5, 6] $\times 10^{18}$ and [7, 9, 10] $\times 10^{18}$ atoms cm$^{-2}$ for the 210\arcsec\ and 105\arcsec\ data 
and constant velocity contours at  [239, 240] \kms.}
\label{fig:hvc271.57+79.03+248}
\end{figure*}

\begin{figure*}
\centering
\includegraphics[keepaspectratio,width=0.6\linewidth]{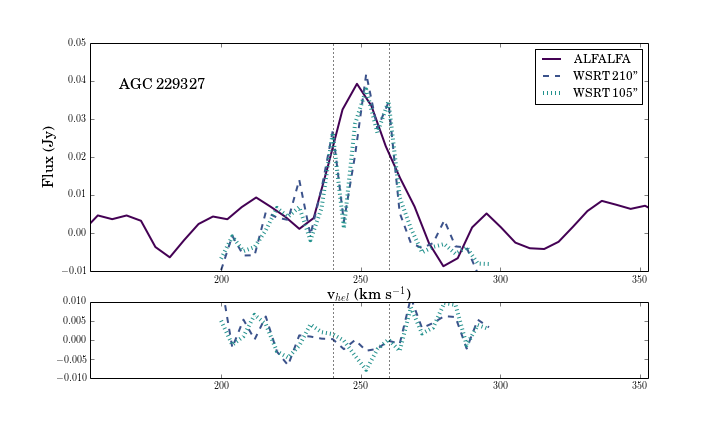}
\includegraphics[keepaspectratio,width=0.35\linewidth]{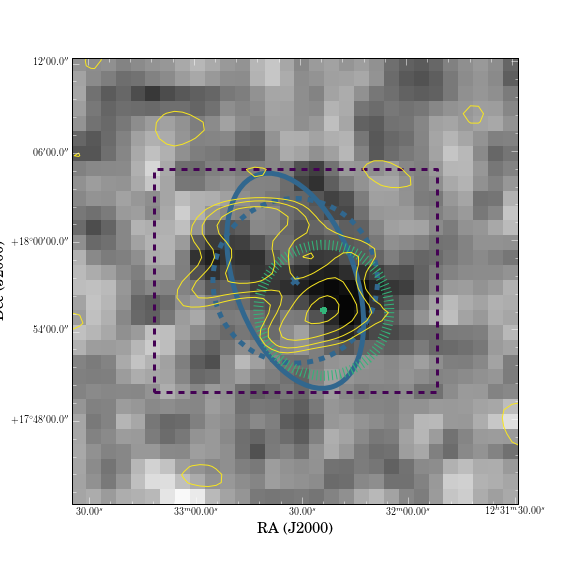}
\includegraphics[keepaspectratio,height=.5\linewidth]{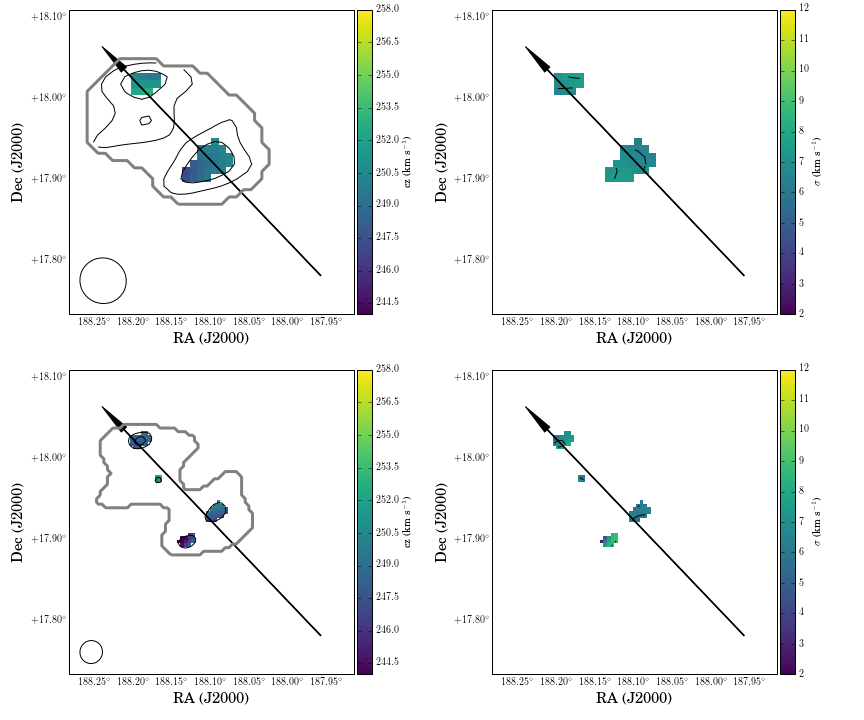}
\includegraphics[keepaspectratio,height=.5\linewidth,clip=true,trim=0cm 2cm 0cm 2.5cm]{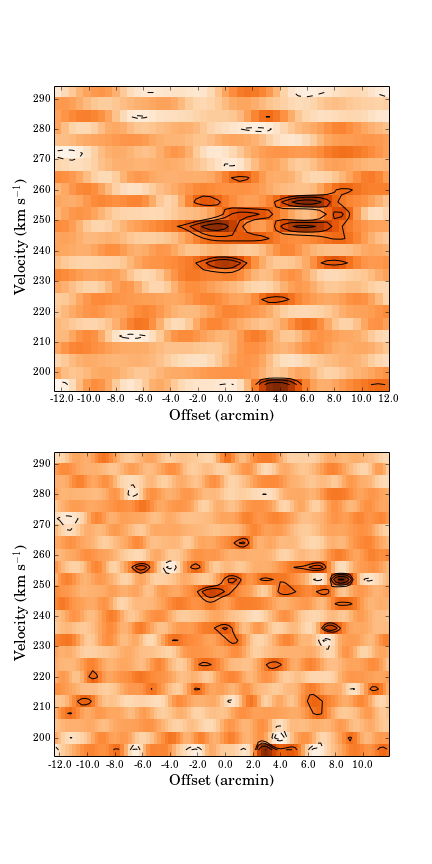}
\caption{ALFALFA and WSRT data for AGC\,229327 as in Figure \ref{fig:hvc214.78+42.45+47} but with \hi\ column density contours at  [2, 3] $\times 10^{18}$ and [6, 8] $\times 10^{18}$ atoms cm$^{-2}$ for the 210\arcsec\ and 105\arcsec\ data and velocity contours at [248, 250, 252] \kms.}
\label{fig:hvc276.53+79.84+255}
\end{figure*}

\begin{figure*}
\centering
\includegraphics[keepaspectratio,width=0.6\linewidth]{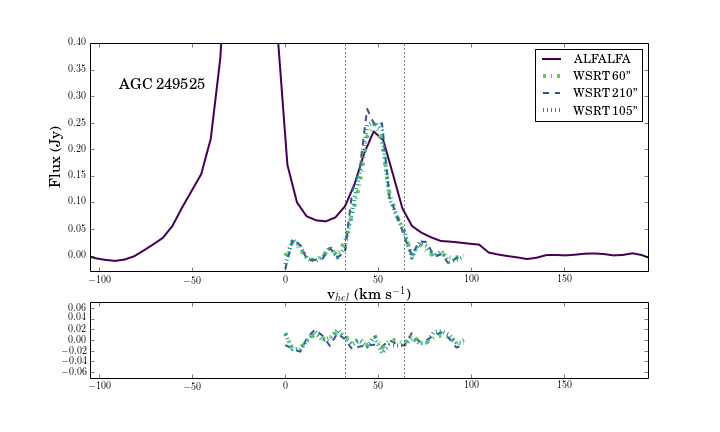}
\includegraphics[keepaspectratio,width=0.35\linewidth]{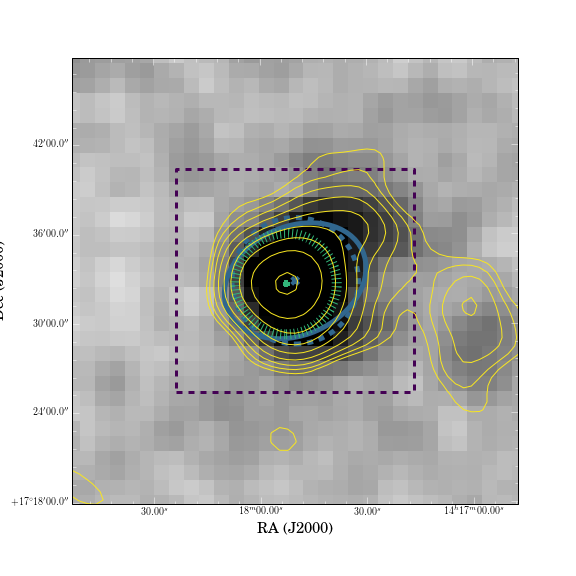}
\includegraphics[keepaspectratio,height=.5\linewidth]{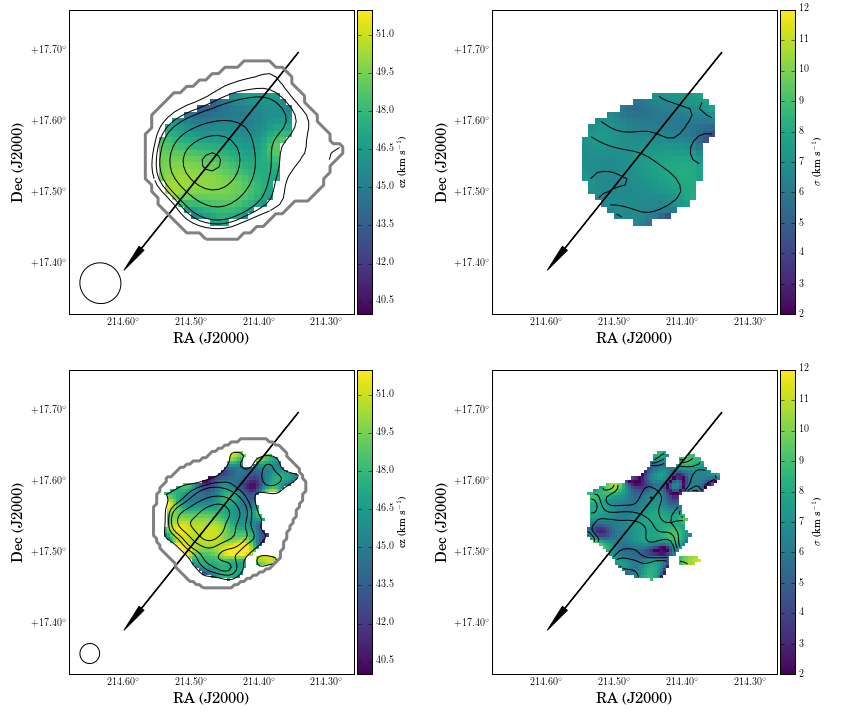}
\includegraphics[keepaspectratio,height=.5\linewidth,clip=true,trim=0cm 2cm 0cm 2.5cm]{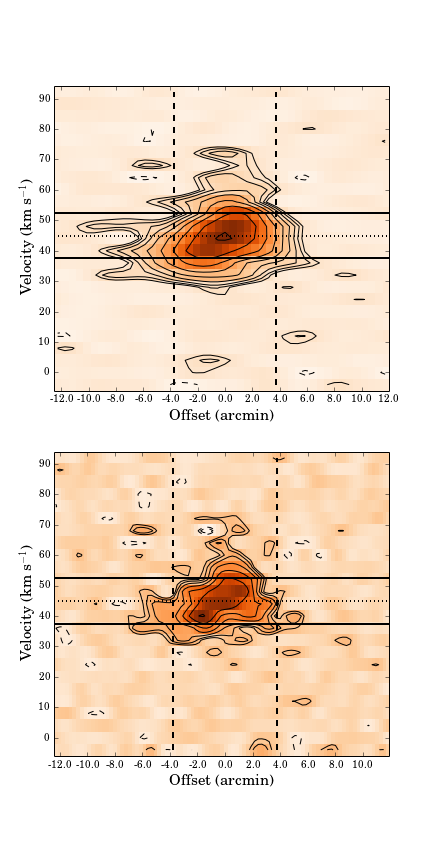}
\includegraphics[keepaspectratio,height=.25\linewidth]{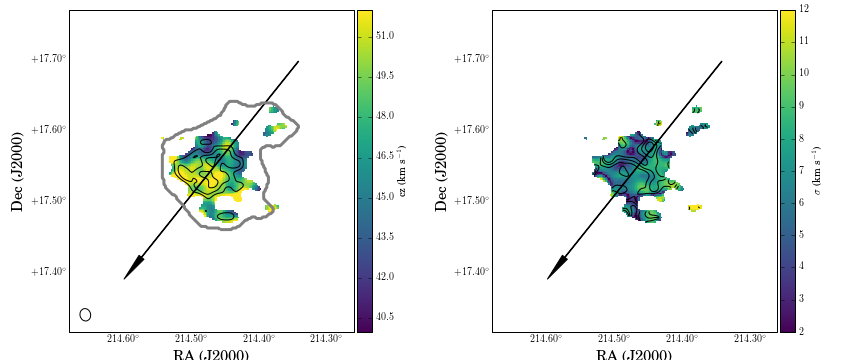}
\includegraphics[keepaspectratio,height=.25\linewidth,clip=true,trim=0cm 1cm 0cm 1.25cm]{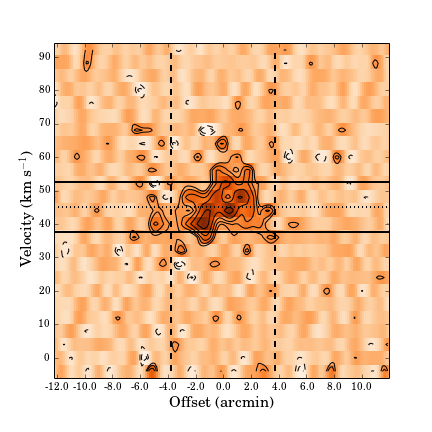}
\caption{ALFALFA and WSRT data for AGC\,249525 as in Figure \ref{fig:hvc214.78+42.45+47} but with \hi\ column density contours at  [4, 6, 8, 15, 25, 35] $\times 10^{18}$, [0.9, 1.5, 2, 3, 4, 5] $\times 10^{19}$, and [3, 4, 5] $\times 10^{19}$ atoms cm$^{-2}$ for the 210\arcsec,105\arcsec, and 60\arcsec\ data and velocity contours at [46, 48, 50] \kms.}
\label{fig:hvc11.76+67.89+60}
\end{figure*}

\begin{figure*}
\centering
\includegraphics[keepaspectratio,width=0.6\linewidth]{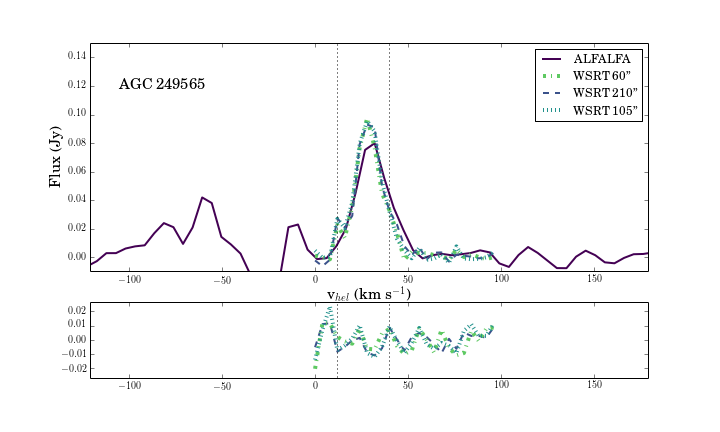}
\includegraphics[keepaspectratio,width=0.35\linewidth]{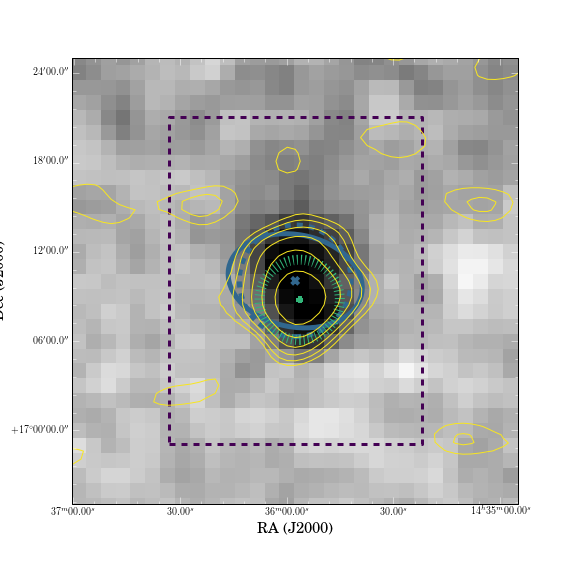}
\includegraphics[keepaspectratio,height=.5\linewidth]{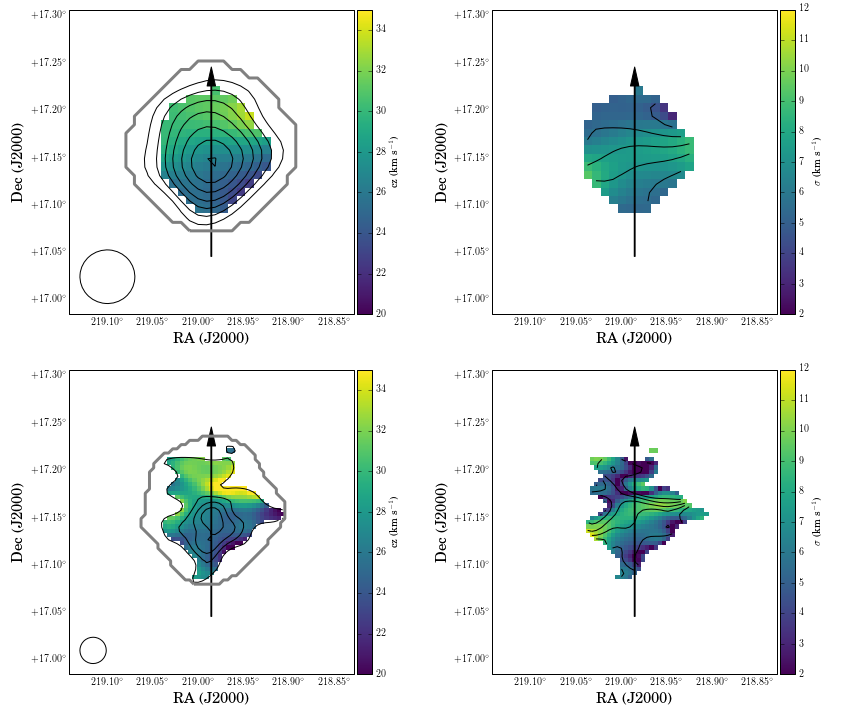}
\includegraphics[keepaspectratio,height=.5\linewidth,clip=true,trim=0cm 2cm 0cm 2.5cm]{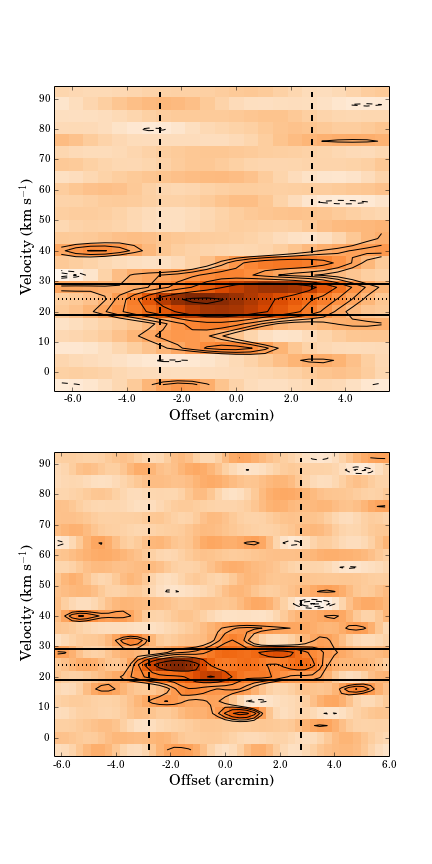}
\includegraphics[keepaspectratio,height=.25\linewidth]{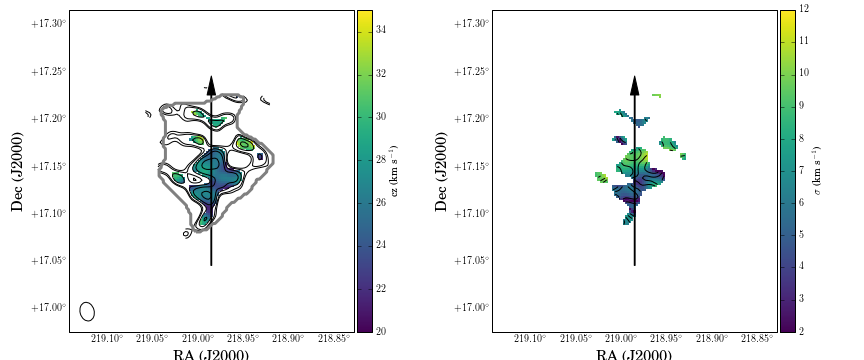}
\includegraphics[keepaspectratio,height=.25\linewidth,clip=true,trim=0cm 1cm 0cm 1.25cm]{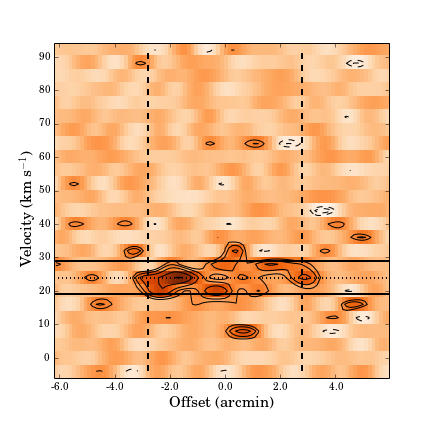}
\caption{ALFALFA and WSRT data for AGC\,249565 as in Figure \ref{fig:hvc214.78+42.45+47} but with \hi\ column density contours at  [4, 6, 8, 10, 12.5, 15, 17.5] $\times 10^{18}$, [0.9, 1.5, 2, 2.5] $\times 10^{19}$,
and  [1.25, 1.5, 2, 2.5, 3.5] $\times 10^{19}$ atoms cm$^{-2}$ for the 210\arcsec, 105\arcsec, and 60\arcsec\ data and velocity contours at [24, 26, 28, 30, 32] \kms.}
\label{fig:hvc15.96+63.90+44}
\end{figure*}

\begin{figure*}
\centering
\includegraphics[keepaspectratio,width=0.6\linewidth]{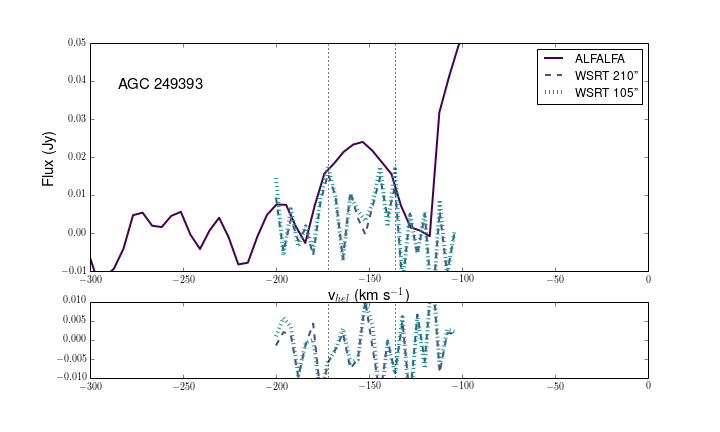}
\includegraphics[keepaspectratio,width=0.35\linewidth]{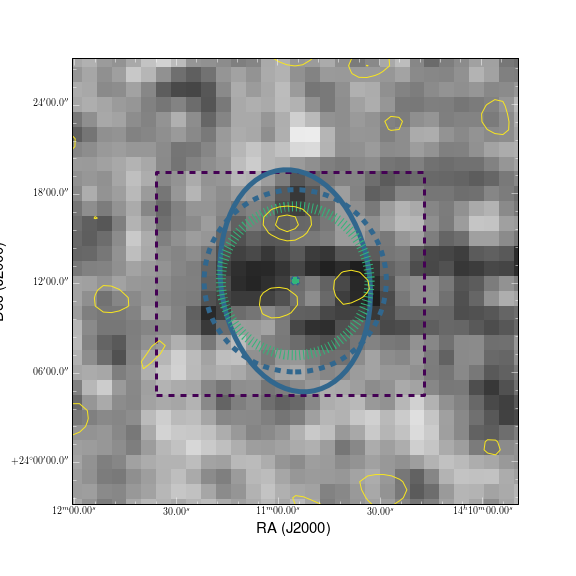}
\caption{ALFALFA and WSRT data for AGC\,249393 as in the top row of Figure \ref{fig:hvc214.78+42.45+47}. 
AGC\,249393 is a non-detection in the WSRT data and
an aperture of radius 5\arcmin\ (shown by the dotted green circle) is used for extracting the spectra.}
\label{fig:hvc28.09+71.87-142}
\end{figure*}

Here we present the ALFALFA and WSRT data for each UCHVC, along with a brief
discussion of the source and results of the WSRT observations.
Figures \ref{fig:hvc214.78+42.45+47}--\ref{fig:hvc28.09+71.87-142} present the WSRT and ALFALFA \hi\ data for each source in a consistent manner.
The left figure in the first row shows the spectra for the primary-beam corrected WSRT data (at all resolutions imaged), 
and the ALFALFA spectrum.
The bottom panel of this figure shows a representative noise spectrum from a signal-free portion of the (non-primary-beam-corrected)
WSRT data cubes.
The right panel in the first row is an ALFALFA grayscale map with the 
(non-primary-beam-corrected) WSRT 210\arcsec\ moment zero map overlaid in units of significance,
and the ALFALFA and WSRT effective half-flux apertures shown.
The following rows present the WSRT data at all relevant angular resolutions.
The left panel is (primary beam corrected)
 \hi\ column density contours  overlaid on the moment one (velocity field) map,
 the center panel is the moment two (velocity dispersion) map with lines of constant velocity overlaid,
 and the right panel is the position-velocity slice, along the direction shown in the two left panels.

{\bf AGC\,198606:} This source was chosen for observation because of its proximity to the gas-rich ultra-faint dwarf galaxy Leo T and similar \hi\ properties within the ALFALFA \hi\ dataset. WSRT observations of this source were also presented in \citetads{2015A&A...573L...3A}. The images here represent a reprocessing of the same data with an additional (heavily flagged) track included.
Figure \ref{fig:hvc214.78+42.45+47} presents the \hi\ data for AGC\,198606.
The source is among the most extended of the UCHVCs observed with WSRT, but it has a smooth \hi\ morphology at both the Arecibo and 105\arcsec\ resolution, as seen in Figure \ref{fig:hvc214.78+42.45+47},
and thus it was also imaged with a 60\arcsec\ taper. 
This source overlaps with the foreground Galactic \hi\ in velocity-space, and
special care is needed to needed to isolate the emission of the source from the foreground.
For example, the ALFALFA source box seen in Figure \ref{fig:hvc214.78+42.45+47}
does not encompass all of the significant emission seen in the WSRT contours;
this is because the presence of the Galactic \hi\ foreground made it 
difficult to determine the physical extent of AGC\,198606  in the ALFALFA data.
The difficultly with the Galactic \hi\ is also seen in the WSRT data;
the apparent turn-over in the velocity gradient seen on the south-eastern edge of  the 210\arcsec\ velocity field is likely
caused by Galactic \hi\ entering the source mask.
As discussed in \citetads{2015A&A...573L...3A}, the velocity fields and position velocity slices of this source at both 210\arcsec\ and 105\arcsec\ resolution show ordered motion with a gradient of $\sim 25$ \kms. 
In this work we see that this velocity motion is also present in the higher angular resolutions 60\arcsec\ data.
This range of velocity motion,
along with the spatial extent at which it is seen at all resolutions, is indicated in the
position-velocity slices in Figures \ref{fig:hvc214.78+42.45+47} and \ref{fig:pvslices}. 
The interpretation of this velocity gradient is discussed in Section
\ref{sec:gal}.
We also note that the integrated flux  derived here is slightly higher than that in \citetads{2015A&A...573L...3A},
although still consistent within the errors. This demonstrates the sensitivity of the final integrated line flux
to the exact source extent.

{\bf AGC\,198511:} 
This source 
is included in the most-isolated subsample of \citetalias{2013ApJ...768...77A}
and was selected for WSRT observations because of its isolation;
it is also one of the most compact sources.
Figure \ref{fig:hvc204.88+44.86+147} presents the \hi\ data for this source.
At the Arecibo resolution, this source has a smooth \hi\ distribution, elongated to the North.
This elongation is also seen in the ALFALFA total intensity \hi\ map.
In addition, both the WSRT and ALFALFA data show that this elongation
connects to another (lower intensity) clump of emission.
At higher angular resolution, 
the main body of the source consists of
separate clumps in an envelope with a morphology reminiscent of 
ram pressure effects. 
 Neither the velocity field nor the position-velocity slice show any strong evidence for ordered velocity motion. 
 Generally this source has a small velocity dispersion. 
The source extent for the WSRT data at 210\arcsec\ resolution includes some of the northern extension,
which was not included in the source box for the ALFALFA data, thus explaining a slightly higher integrated flux density for 
the WSRT 210\arcsec\ data than the ALFALFA data.

{\bf AGC\,198683:}
This source was also selected for its isolation; it is relatively close to AGC\,198511 but it is larger
and has a lower column density in the ALFALFA \hi\ data, making it a less promising candidate. 
The \hi\ data for this source are presented in Figure \ref{fig:hvc205.83+45.14+173}.
The WSRT data quality for this source is relatively poor, as can be seen in the noisy spectra.
At the 210\arcsec\ resolution, the source shows elongation to the north, which is aligned
with the extension seen in the ALFALFA grayscale map.
This structure is also reflected in the ALFALFA half-power ellipse.
 At higher angular resolution, 
 the main body of source emission resolves into two  clumps while the northern clump
 is also detected. 
 Only a fraction of the ALFALFA flux is detected, indicating that much of the emission 
 is low surface brightness and extends further than the flux masks indicate.
 This is  consistent with less flux being recovered in the 105\arcsec\ data which is less
 sensitive to low column density emission.
 The source has no velocity structure.

{\bf AGC\,208747:}
This source was selected for having a high average column density value from the ALFALFA data.
Figure \ref{fig:hvc217.77+58.67+96} presents the \hi\ data for AGC\,208747.
 At ALFALFA resolutions, the source is smooth and slightly elongated. 
 At higher angular resolutions, the main body shows some clumpiness but is generally smooth.
  The tail separates into a separate clump of emission, connected by low level emission to the main body.
  The velocity fields appear to indicate a potential velocity gradient across the source,
   but there is no evidence for ordered velocity motion
  in the position-velocity slices.
  The WSRT data match the blue end of the ALFALFA spectrum well but are missing emission on the red end.

{\bf AGC\,208753:}
This source was chosen based on isolation and a high average column density value from the ALFALFA data. 
The \hi\ data for AGC\,208753 are presented in Figure \ref{fig:hvc212.68+62.39+64}.
At the ALFALFA resolution it is elongated in the north-south direction and consists of two separate clumps connected at the level of $5 \times 10^{18}$
atoms cm$^{-2}$. This structure is tentatively seen in the ALFALFA grayscale map, although
 the position angle of the ALFALFA half-power ellipse is perpendicular to the elongation.
At higher angular resolution, this source breaks into multiple clumps, in a common envelope around $5 \times 10^{18}$
atoms cm$^{-2}$. 
No velocity structure is evident in the velocity field or position velocity slices. 
The clumpy nature of this source at higher angular resolution is also seen in the position-velocity slice.
The ALFALFA flux is well recovered by both the 210\arcsec\ and 105\arcsec\ data.

{\bf AGC\,219663:} 
This object was selected for its compact size. 
Figure \ref{fig:hvc230.27+71.10+76} presents the \hi\ data for AGC\,219663.
At the ALFALFA resolution, this source shows very little structure but does have 
a hint of elongation to the north-east.
This elongation is also seen in the ALFALFA grayscale map and in the position and orientation of the
ALFALFA half-power ellipse.
The source has very little velocity structure; there is an apparent gradient in the velocity field but it is only of
order $\sim 5$ \kms, and 
 in the position-velocity diagram there appears to be two separate velocity structures.
The source has low velocity dispersions across the entire extent of emission.
At higher angular resolution, the source appears as two clumps that are barely detected.
These appear to be the two separate velocity structures hinted at in the 210\arcsec\ data; this is
also seen in the position-velocity slice.
The total integrated flux density recovered is slightly lower than the ALFALFA value, although 
the emission detected in the WSRT cubes appears at the 210\arcsec\ resolution appears to be
narrower in velocity width than that of the ALFALFA data, as seen in the spectra figure.

{\bf AGC\,2196565:}
This UCHVC was selected for its fairly compact \hi\ size and isolation in that ALFALFA data. 
The \hi\ data are shown in Figure \ref{fig:hvc235.38+74.79+195}.
At the ALFALFA resolution, it is extended in the northeast-southwest direction; evidence for this elongation is also seen in the ALFALFA map.
The WSRT data also show a secondary structure, perpendicular to the main axis. 
This is not obviously evident in the ALFALFA data.
At higher angular resolution, the source breaks into multiple clumps along the axis of elongation,
and there are are also two separate clumps associated with the perpendicular structure. 
There is no clear velocity structure in the source, in either the velocity fields or position-velocity slices. 
We note that this source has a higher measured integrated flux density from the WSRT data than the ALFALFA data.
There is no clear explanation for this as the ALFALFA source box encompasses all of the WSRT emission 
and this object is well removed from the Galactic foreground in velocity space.
The perpendicular structure is not seen in the ALFALFA data and the inclusion of it in the WSRT maps
may account for the larger recovered flux.
As can been seen in the bottom of the spectra panel in Figure \ref{fig:hvc235.38+74.79+195}, 
the noise levels for this source are high, and it is possible that this extra structure
is a left-over imaging artifact.

{\bf AGC\,229326:}
This source was selected for observation because of its relatively high recessional velocity, typically difficult to explain in Galactic fountain models of high velocity clouds. 
Figure \ref{fig:hvc271.57+79.03+248} presents the WSRT and ALFALFA \hi\ data. 
At the Arecibo resolution, this source is compact with a slightly extended \hi\ tail.
 At higher angular resolutions, the source breaks into three distinct clumps,
 with one of these in the tail seen at lower resolution. 
As seen in the position-velocity slices, 
AGC\,229326 appears to consist of clumps of emission with narrow velocity widths.
The WSRT spectra lack emission on the red end compared to the ALFALFA spectrum.
 This is a low S/N detection in the WSRT data, as evident in the noise in both the spectra and position-velocity slices.

{\bf AGC\,229327:}
This object was chosen for its high recessional velocity, although it is also among the largest and lowest surface brightness of the ALFALFA UCHVCs targeted for follow-up. 
Figure \ref{fig:hvc271.57+79.03+248} presents the \hi\ data.
At 210\arcsec\ resolution it appears as two clumps; this is seen also in the ALFALFA data.
The ALFALFA half-power ellipse does a good job of matching the elongation of the system due to the two components,
but the position angle is mis-aligned.
At the 105\arcsec\ resolution, two cores in each clump are apparent. 
These cores have very narrow velocity extents as seen in the position-velocity slices.
 There is no clear velocity structure in the velocity fields but the position-velocity slices, especially at 210\arcsec\ resolution, show
 evidence for multiple velocity features at the same location.
Similar spectra and total integrated flux densities are found for both the 210\arcsec\ and 105\arcsec\ WSRT 
data; however both of these spectra are low S/N compared to the ALFALFA data.

{\bf AGC\,249525:}
This UCHVCs was selected for its high $\bar{N}_{HI}$. 
The \hi\ data are shown in Figure \ref{fig:hvc11.76+67.89+60}.
At 210\arcsec\ resolution it is very smooth with some hints of extension and evidence for a velocity gradient. 
At the 105\arcsec\ resolution,
the source shows a more pronounced velocity gradient and still has a very smooth \hi\ morphology,
although the outer \hi\ contour is slightly disturbed.
Since the source is still strongly detected at 105\arcsec\ resolution,
it was also imaged with a 60\arcsec\ taper.
At that resolution, it is still detected and shows the velocity gradient.
The extent of this velocity gradient is $\sim 15$ \kms, as indicated by the solid lines in the position-velocity 
slices (around a central velocity of 45 \kms) in Figures \ref{fig:pvslices} and \ref{fig:hvc11.76+67.89+60}. 
This velocity gradient persists across $\sim$7.5\arcmin, also indicated in the position-velocity slices.
The implication of this velocity gradient is discussed in Section \ref{sec:gal}.
At the low velocity end, the emission flattens to a constant velocity. 
This is where the elongated tail of emission is located.
The clump of emission seen to the west in both the ALFALFA grayscale map and the WSRT data
is likely Galactic foreground emission; it was excluded from the mask encompassing the source.
The spectra of this source are in good agreement with the ALFALFA data except for a slight lack
of emission at the red end.

{\bf AGC\,249565:}
This source was chosen for its high $\bar{N}_{HI}$. 
Figure \ref{fig:hvc11.76+67.89+60} shows the \hi\ data for AGC\,249565.
At the ALFALFA resolution, its \hi\ morphology is exceptionally smooth
and it shows evidence for ordered motion in its velocity field (although it is barely resolved).
At the higher 105\arcsec\ angular resolution, it shows some elongation, especially in the outer \hi\ contours, but the core of the source remains smooth and undisturbed, and the velocity gradient is still present
 in the velocity field and position-velocity slice, although the position-velocity slice
also shows some clumpy structure.
Since this source was robustly detected in the 105\arcsec\ WSRT data, it was also imaged with a 60\arcsec\ taper.
At this resolution its \hi\ morphology starts to become clumpy, but this may be a S/N limitation rather than
a true change in morphology.
In the position velocity slices in Figures \ref{fig:pvslices} and \ref{fig:hvc11.76+67.89+60}, 
a velocity extent of 10 \kms\ is indicated by the solid lines (around a center
of 24 \kms), with a spatial extent of 5.6\arcmin\ indicated by the vertical dashed line.
The velocity extent is most clearly seen in the 210\arcsec\ data, although it is still present in the
outer envelope at higher resolution.
The implication of this is discussed in Section \ref{sec:gal}.
The WSRT spectrum matches the ALFALFA spectrum well.

{\bf AGC\,249393:}
This object was selected for its isolation within the ALFALFA survey data. It is the largest and lowest $\bar{N}_{HI}$ source observed with WSRT. Unfortunately, it is also the only source for which the final imaging was done with only a single track. (The second track had a substantial number of antennas missing and so could not be effectively tapered.)
The source is an apparent non-detection, but is consistent with being at the noise level. In Figure \ref{fig:hvc28.09+71.87-142}
the ALFALFA spectrum is 
shown with spectra extracted from the WSRT dirty cubes with 
an aperture of radius 5\arcmin, consistent with the ALFALFA \hi\ size. The noise in the WSRT spectra are consistent with the non-detection. In addition,
confidence contours from a moment zero map of the dirty data cube (velocity range indicated in the spectra figure) are shown, and there is evidence at the 2-$\sigma$ level that the source is present. However, we were unable to isolate the source for cleaning and do not report any \hi\ parameters from the WSRT data except for an approximate peak column density value based on the moment zero map from the dirty data. The peak column density seen in the 210\arcsec\ map is consistent with the ALFALFA $\bar N_{HI}$ value.

\end{document}